\documentclass[superscriptaddress,
amsmath,amssymb,
aps,
pra,
floatfix,
twocolumn]{revtex4-2}
\usepackage{graphicx} 
\usepackage{float}
\usepackage{braket}
\usepackage{bm}
\usepackage{amsmath}
\usepackage{amssymb}
\usepackage{placeins}
\usepackage{hyperref}
\usepackage{siunitx}
\usepackage{comment}
\begin{document}

\newcommand{\asc}[1]{\textcolor{red}{#1}}
\newcommand{\jdw}[1]{\textcolor{blue}{JDW: #1}}
\newcommand{\DD}[1]{\textcolor{green}{#1}}
\title{Quantum-enhanced single and multiparameter metrology in qutrit ensembles by generalized twisting dynamics}

\newcommand{\iitm}{\affiliation{Department of Physics, Indian Institute of Technology Madras, Chennai 600036, India.}}
\newcommand{\cquicc}{\affiliation{ Center for Quantum Information, Communication and Computing, Indian Institute of Technology Madras, Chennai 600036, India}}
\newcommand{\jila}{\affiliation{Department of Physics, University of Colorado, Boulder, CO 80309}}

\author{Deep Datta}
\email{ph24d007@smail.iitm.ac.in}
\iitm
\cquicc
\author{Sayam Chakraborty}
\iitm
\cquicc
\author{John Drew Wilson}
\jila
\author{Vaibhav Madhok}
\iitm
\cquicc
\author{Athreya Shankar}
\email{athreya@physics.iitm.ac.in}
\iitm
\cquicc

\begin{abstract}
    Quantum-enhanced sensing with atomic ensembles has predominantly focused on qubit-based protocols, despite the growing ability of many experimental platforms to coherently control and entangle multi-level systems. Here, we investigate quantum-enhanced sensing with qutrit ensembles by introducing three experimentally feasible qutrit generalisations of the one-axis twisting (OAT) model that involve entangling operations only between two levels, while the third level primarily acts as a spectator. We characterize the metrological utility of the dynamically generated states using the quantum Fisher information toolbox. We find numerically that all three models offer considerable freedom in encoding direction for quantum-enhanced sensing, with up to 6 out of 8 possible directions exhibit near-Heisenberg scaling after a short evolution time. We discuss experimental access to this enhanced metrological precision via effective time-reversal protocols. Furthermore, we examine the practical issues of estimation ambiguity and local dissipation, and show that they can be largely overcome by optimizing the sensor operating point. Finally, we show that the measurement incompatibility in estimating multiple parameters simultaneously encoded in different directions with near-Heisenberg scaling of precision is suppressed at the zero operating point as the system size increases. In the process, we find that one of the models enables near-Heisenberg scaling metrology with a pair of commuting generators, a possibility that arises from the \(su(3)\) algebra and is thus absent in qubit-ensemble based sensors of collective $SU(2)$ rotations.    
\end{abstract}

\date{\today}
\maketitle

\section{Introduction}
Atomic ensembles have proven instrumental in quantum sensing and metrology due to their long coherence times, high controllability, and strong coupling to external signals such as electric and magnetic fields and gravity~\cite{degen2017quantum,pezze2018quantum,montenegro2025quantum}. Such high-precision measurements are essential for multiple applications, ranging from navigation to clocks for precise timekeeping to highly sensitive magnetometry, and also for searches of new physics~\cite{ludlow2015optical,kasevich1991atomic,budker2007optical,bongs2019taking}. In practice, the precision of these measurements is often limited by the projection noise of independent atoms, which sets a lower bound on the achievable precision with uncorrelated atoms, typically called the standard quantum limit (SQL).
Quantum-enhanced sensing aims to push precision below the SQL by utilising entangled states and tailored measurements, and ultimately achieve the so-called Heisenberg limit(HL), which represents the optimal precision attainable for linear quantum metrology protocols~\cite{giovannetti2004quantum,pezze2018quantum}. Several works have explored protocols for quantum-enhanced metrology with ensembles of two-level atoms or qubits. For example, spin squeezing protocols involving non-linear two-body interactions between qubits have been proposed and experimentally realised on several platforms~\cite{kitagawa1993squeezed,riedel2010atom,sewell2012magnetic,hosten2016measurement}. However, recent experimental progress on several platforms such as trapped ions and neutral atoms has made it possible to coherently control more than just two atomic levels ~\cite{hrmo2023native,burshtein2026robust,shlyakhov2018quantum,scazza2014observation}. These advancements raise the question of how we can exploit qudit ensembles for quantum-enhanced sensing and whether the enlarged Hilbert space leads to new features that are beneficial for quantum sensor operation. 

In this work, we take a step in this direction by introducing and characterising protocols for quantum-enhanced metrology with ensembles of three-level atoms or qutrits. In particular, we consider three extensions of the paradigmatic one-axis twisting (OAT) Hamiltonian~\cite{kitagawa1993squeezed} to qutrits and characterise the metrological utility of the resulting states. A central finding is that all the considered models offer considerable freedom in signal encoding, with up to $6$ out of $8$ possible encoding directions showing a precision that scales approximately like the Heisenberg limit, up to a constant factor. Furthermore, the enhanced sensitivities are experimentally accessible via a time-reversal protocol, which is compatible with the OAT Hamiltonian and has been previously realized~\cite{davis2016approaching,colombo2022time}. To circumvent the ambiguity in estimating the phase at the zero operating point, we propose operating at a non-zero sensor operating point. We also examine the impact of dissipation on the sensor operation and find that its impact can be largely mitigated by optimizing the sensor operating point. Finally, we explore the possibility of compatible multiparameter estimation by simultaneously encoding parameters in multiple generators and find a new route to compatibility that is absent in qubit ensembles.

Our work complements recent experimental progress in controlling and manipulating spin-1 Bose-Einstein condensates (BEC) for quantum sensing applications. For instance, nematic spin squeezing of a spin-1 atomic BEC in the \(F=1\) ground state has been combined with an echo protocol to achieve quantum-enhanced sensing~\cite{You2023quantum}. Spin-1 Dicke states have also been experimentally generated by driving a BEC through a quantum phase transition,  and the resulting states have been used in interferometry to beat classical sensing limits~\cite{You2018beating}. Our work introduces approaches to qutrit sensors that are broadly applicable on platforms that support all-to-all coupling, such as trapped ions and atom-cavity systems. Importantly, the qutrit extensions we propose require only entangling operations in a two-level subspace, with the third level serving as a spectator during the entangling dynamics. This means they can be readily implemented on any platform where standard OAT is available, and an additional third level can be coherently manipulated using global single-particle drives.

This paper is organised as follows. In Sec.~\ref{review_QM}, we provide a brief review of the concepts of quantum metrology relevant to this work, covering both single-parameter and multiparameter estimation. In Sec.~\ref {theoretical model}, we introduce the three qutrit extensions of the OAT Hamiltonian considered in this study along with their experimental realisation in both trapped ion and cavity setups. The metrological utility of the resulting states for measuring a single unknown parameter is characterised in terms of the quantum Fisher information (QFI) in Sec.~\ref{single_para}, and experimental protocols to saturate the QFI are discussed in Sec.~\ref{attainibility_QFI}. In Sec.~\ref {prac_con}, we discuss some practical considerations of this protocol. In particular, in Sec.~\ref {unambiguous_phase}, we circumvent the ambiguity in phase estimation near the zero operating point, while in Sec .~\ref {Dephasing}, we study the performance of the time-reversal protocol under local decay and dephasing and determine optimal sensor operating points to mitigate their effects. Subsequently, in Sec.~\ref{MUL}, we examine the compatibility of simultaneous precision estimation of multiple parameters. We conclude with a summary and outlook in Sec.~\ref{sec:conc}.

\section{Review of Quantum metrology}
\label{review_QM}
In this section, we review the main concepts of quantum metrology relevant to this work. We first discuss the setting for single-parameter metrology, introduce the quantum Fisher information (QFI) and related bounds, and discuss optimal generators for parameter encoding. We then briefly review the setting for multiparameter metrology, introduce the QFI matrix and discuss the issue of compatibility. 

\subsection{Single parameter estimation}
\label{single_para_theory}
We consider a typical quantum metrology protocol corresponding to unitary phase estimation. Here, the protocol can be divided into four parts: probe state preparation, unitary parameter encoding, measurement in a chosen basis and estimation of the unknown parameter from the measurement data. Let $\rho$ be the prepared probe state. The unknown parameter $\theta$ is encoded in the state via a unitary transformation that modifies the probe state to
\begin{equation}
\label{state_encoding}
    \rho_{\theta}=e^{-i\theta{G}}\rho e^{i\theta{G}}.
\end{equation}
Here, the parameter $\theta$ is encoded into the probe state via a unitary transformation effected by the Hermitian operator $G$ or the generator. Subsequently, a measurement is performed in a chosen basis, leading to outcomes $x$ with associated probabilities $p(x|\theta)$. Here, we assume the measurements to be rank-$1$ projective measurements, corresponding to projections on an orthonormal complete basis $\{\ket{m}\}$. In this case,
\begin{equation}
    p(x|\theta) = \braket{m|\rho_{\theta}|m}.
\end{equation}

In general, the entire protocol may be repeated $M$ times to accumulate measurement statistics, leading to a record of measurement outcomes $\boldsymbol{x}=(x_1,\ldots,x_{M})^T$.  Finally, a mapping function called the estimator maps the observed measurement record to an estimate of the unknown phase. We denote the estimator by $\theta_{\rm est}(\boldsymbol{x})$. 

The figure of merit for the above measurement protocol is the mean squared error (MSE)  averaged over all measurement records, given by
\begin{equation}
    \epsilon = \sum_{\boldsymbol{x}} (\theta_{\rm est}(\boldsymbol{x}) - \theta)^2 p(\boldsymbol{x}|\theta).
\end{equation}
As each repetition of the experiment is independent, $p(\boldsymbol{x}|\theta)=p(x_1|\theta)p(x_2|\theta)\ldots p(x_M|\theta)$. For an unbiased estimator, the MSE for a fixed input state and measurement basis is lower bounded as~\cite{braunstein1994statistical,paris2009quantum}
\begin{equation}
\epsilon \ge \frac{1}{M F_C(\theta)},
\end{equation}
where $F_C(\theta)$ is the classical Fisher information (CFI) given by 
\begin{equation}
F_C(\theta)=\sum_x \frac{1}{p(x|\theta)}\left(\frac{\partial p(x|\theta)}{\partial \theta}\right)^2.
\end{equation}
Here, the sum over $x$ runs over all possible measurement outcomes in a single repetition of the experiment. The CFI is generally dependent on the true parameter value $\theta$, which we refer to as the sensor's operating point. Saturation of the lower bound requires the use of an optimal estimator to map the outcome to the unknown parameter. In general, the optimal estimator may depend on the true parameter value, which makes its construction challenging. However, in the limit of a large number of measurement repetitions, the maximum likelihood estimator is asymptotically optimal under fairly general conditions of regularity and ~\cite{steven1993fundamentals,paris2009quantum}. 

Quantum mechanics allows us freedom in the choice of measurement basis, which can be exploited to reduce the MSE. Maximising the CFI over all possible measurement bases leads to the quantum Fisher information (QFI). If the probe state is pure, i.e., $\rho=\ket{\psi}\bra{\psi}$, the QFI is independent of the operating point and is given by~\cite{braunstein1994statistical,paris2009quantum},
\begin{equation}
    F_Q = \max_{\{\ket{m}\}} F_C(\theta) = 4(\Delta G)^2.
\end{equation}
In other words, the QFI is just four times the variance of the generator used to encode the unknown parameter in Eq.~(\ref{state_encoding}) with respect to the probe state $\ket{\psi}$. Using an optimal measurement basis, the MSE is lower bounded as ~\cite{braunstein1994statistical,paris2009quantum}
\begin{equation}
    \epsilon \geq \frac{1}{M F_C(\theta)} \geq \frac{1}{M F_Q}.
\end{equation}

 Finally, the probe state provided to the sensor can also be optimised to maximise the QFI. Here, it is useful to distinguish initial product states from entangled states. In typical metrology settings, the phase is independently encoded in the sensor atoms via their interaction with a background field of interest, leading to $G=\sum_j g^{(j)}$, where $g^{j}$ is a single-particle operator encoding the parameter in atom $j$. Fixing the normalisation of the operators such that ${\rm Tr}[g^{(j)2}]=1$ and preparing each atom in an equal superposition of the extremal eigenvectors of $g$ leads to the so-called standard quantum limit~\cite{giovannetti2004quantum,giovannetti2011advances,pezze2018quantum}, given by 
\begin{equation}
\label{SQL}
    F_Q \leq N (\lambda_{\rm max}-\lambda_{\rm min})^2,
\end{equation}
where $\lambda_{\rm max}$ and $\lambda_{\rm min}$ are the maximum and minimum eigenvalues of $g$. On the other hand, preparing the probe state in an equal superposition of the extremal eigenvectors of the collective operator $G$ results in an entangled state that attains the Heisenberg limit~\cite{giovannetti2004quantum,giovannetti2011advances,pezze2018quantum}, given by 
\begin{equation}
\label{HL}
    F_Q \leq N^2 (\lambda_{\rm max}-\lambda_{\rm min})^2.
\end{equation}

In this work, we are primarily interested in the dynamical generation of entangled probe states with QFI scaling as $N^2$, although the prefactor may be smaller than the one above. Sensors with $N^2$ scaling of the QFI are said to exhibit Heisenberg \emph{scaling}~\cite{pezze2018quantum} (as opposed to \emph{limit}) of the measurement precision. At this point, one may note that choosing appropriate normalisation in Eq.~(\ref{SQL}) and Eq.~(\ref{HL}) fixes the overall scaling of the QFI such that \(F_Q=N\) corresponds to the standard quantum limit and \(F_Q=N^2\) corresponds to the Heisenberg limit.

\subsubsection{Optimal generators}
To extract the maximum metrological utility of a prepared probe state, it is useful to determine the optimal generator for which the quantum Fisher information (QFI) is maximized~\cite{reilly2023optimal}. An arbitrary generator can be expressed as
\begin{equation}
    G(\boldsymbol{n})=\sum_k n_k G_k,
\end{equation}
where \(\boldsymbol{n}\) is a real unit vector satisfying \(\sum_k n_k^2=1\), and \(\{G_k\}\) is a set of Hermitian generators, which are orthogonal and normalized according to
\begin{equation}
    {\rm Tr}(G_i G_j)=C\delta_{ij},
\end{equation}
with \(C\) a normalization constant. The QFI associated with the generator \(G(\boldsymbol n)\) is
\begin{equation}
    F_Q=\boldsymbol{n}^T \Sigma \boldsymbol{n},
\end{equation}
where \(\Sigma\) is the covariance matrix with elements
\begin{equation}
\label{QFI}
    \Sigma_{ij}=4\,{\rm Cov}(G_i,G_j).
\end{equation}
The optimal generator is determined by the eigenvector \(\boldsymbol{n}_{\rm opt}\) of \(\Sigma\) corresponding to the largest eigenvalue \(\lambda_{\rm max}\), namely
\begin{equation}
\label{G_opt_gen}
    G_{\rm opt}=\sum_k (\boldsymbol{n}_{\rm opt})_k G_k.
\end{equation}

\subsection{Multiparameter estimation and Compatibility}
\label{multi_comp}
An interesting question is whether, instead of only encoding along the optimal direction, one can simultaneously take advantage of the other sensitive directions (eigen generators of QFIM) by using two or more such directions to encode a vector of unknown parameters \(\boldsymbol{\theta}=(\theta_1,\theta_2,\ldots)\).
The total mean-squared estimation error is then given by
\begin{equation}
    \epsilon_m
    =
    \sum_{\boldsymbol{x}}
    \left\|
    \boldsymbol{\theta}_{\rm est}(\boldsymbol{x})
    -
    \boldsymbol{\theta}
    \right\|^2
    p(\boldsymbol{x}|\boldsymbol{\theta}),
\end{equation}
where \(p(\boldsymbol{x}|\boldsymbol{\theta})\) is the probability
distribution of measurement outcomes \(\boldsymbol{x}\). 
The single-parameter quantum Cramer-Rao bound can then be naturally extended to multiple parameters as~\cite{szczykulska2016multi,liu2020quantum}
\begin{equation}
\label{MQCRB}
    \epsilon_m\geq \frac{\mathrm{Tr}( \boldsymbol{F_Q}^{-1})}{M}.
\end{equation}
Here \(\boldsymbol{F_Q}\) is the Quantum Fisher Information Matrix (QFIM). Under unitary encoding using \(U=e^{-i\sum_j\theta_jG_j}\) and at the parameter value (operating point) $\boldsymbol{\theta}=\vec{0}$, the QFIM is the same as the covariance matrix \(\Sigma\). The elements of QFIM are then given by~\cite{braunstein1994statistical,pezze2018quantum},
\begin{equation}
\label{COV_mat}
    \boldsymbol{[F_Q]_{ij}}=\Sigma_{ij}=4\,{\rm Cov}(G_i,G_j).
\end{equation}
In this work, we will focus only on the zero operating point $\boldsymbol{\theta}=\vec{0}$ when we study multiparameter metrology, unless otherwise mentioned. 

Unlike the single-parameter case, however, the multiparameter quantum Cramer-Rao bound is not always saturable using a single measurement basis. For pure states, the limitation arises from the non-commutation between the generators used to encode the parameters. Saturability can be achieved when the encoded parameters are compatible~\cite{ragy2016compatibility}. For simultaneous estimation of multiple unknown parameters, the mean Uhlmann curvature \(U_{\mu\nu}\) can be used to characterise the compatibility of the unknown parameters. For a pure state, the mean Uhlmann Curvature is given by,
\begin{equation}
\label{UC}
    U_{\mu\nu}=-\frac{i}{4}\langle[G_{\mu},G_{\nu}]\rangle.
\end{equation}
The Uhlmann curvature is, in general, expressed in terms of the commutators of the symmetric logarithmic derivatives associated with each parameter~\cite{ragy2016compatibility}. However, for unitary encoding and at the zero operating point, it can be expressed directly in terms of the encoding generators as written in Eq.~(\ref{UC}).

Physically, \(U_{\mu\nu}\) captures the extent to which the generators fail to commute with the support of the state. The saturability of the multi-parameter quantum Cramer-Rao bound is guaranteed when the mean Uhlmann curvature vanishes~\cite{carollo2019quantumness,ragy2016compatibility,yamagata2013quantum}, i.e.,
\begin{equation}
    U_{\mu\nu}=0, \quad \forall \; \mu,\nu .
\end{equation}

Physically, this condition implies that the statistical model does not exhibit a geometric obstruction to the simultaneous estimation of the parameters using a single measurement basis.

In general, due to incompatibility, a stricter bound called the Holevo Cramer-Rao bound~\cite{holevo2011probabilistic} denoted by $C_H$ is saturable, leading to the hierarchy of inequalities
\begin{equation}
    \epsilon_m\geq\frac{C_H}{M}\geq \frac{\mathrm{Tr}( \boldsymbol{F_Q}^{-1})}{M}.
\end{equation}
While $C_H$ is in general challenging to compute, Ref.~\cite{razavian2020quantumness} showed that  
\begin{equation}
    \mathrm{Tr}( \boldsymbol{F_Q}^{-1})\leq C_H\leq (1+R)\mathrm{Tr}( \boldsymbol{F_Q}^{-1}),
    \label{eqn:hcrb_bounds}
\end{equation}
where the incompatibility is quantified by the \(R\) parameter, which has a simple closed-form expression given by 
\begin{equation}
\label{R_para}
    R=||2 i \boldsymbol{F_Q}^{-1}U||_{\infty}.
\end{equation}
Here, \(||O||_{\infty}\) denotes the largest eigenvalue of the matrix \(O\). Furthermore, it can be shown that \(0\leq R\leq 1 \) where \(R=0\) corresponds to exact compatibility and \(R=1\) refers to maximum incompatibility. Thus, \(R=1\) gives an upper bound to the \(C_H\), i.e., 
\begin{equation}
    C_H\leq 2\, \mathrm{Tr}{(\boldsymbol{F_Q}^{-1})}.
\end{equation}

\section{Qutrit extensions of OAT}
\label{theoretical model}
In this section, we introduce the theoretical models studied in this work. We begin by introducing the collective operators required to describe qutrit dynamics. Then we review the standard one-axis twisting (OAT) model, which serves as the foundation for our extension to qutrit systems. Building upon the collective qutrit operators, we construct and analyse three qutrit models: the qutrit one-axis twisting (QOAT), the balanced one-axis twisting (BOAT), and the qutrit XY (QXY) model.
\subsection{Collective operators for qutrit ensembles}
\label{Collective}
We consider $N$ three-level systems (qutrits), each with three levels $\ket{0},\ket{1}$ and $\ket{2}$. The collective coherent dynamics of qutrits form a representation of the group \(\mathrm{SU}(3)\). The generator of these dynamics forms the algebra \(\mathrm{su(3)}\). For a single qutrit, the corresponding Lie algebra \(\mathrm{su}(3)\) is spanned by the eight Gell-Mann matrices \(\{\lambda_k\}_{k=1}^{8}\)~\cite{gell2018eightfold}, which therefore constitute an operator basis for \(\mathrm{su}(3)\). These Hermitian matrices are given by 
\begin{align}
\label{Gell-Mann}
\lambda_1 &= (|0\rangle\langle1| + |1\rangle\langle0|), \; \lambda_2 =-i(|0\rangle\langle1| - |1\rangle\langle0|),  \nonumber\\
\lambda_4 &= (|0\rangle\langle2| + |2\rangle\langle0|), \; \lambda_5 = (-i|0\rangle\langle2| + i|2\rangle\langle0|),\nonumber\\
\lambda_6 &= (|1\rangle\langle2| + |2\rangle\langle1|), \; \lambda_7 = (-i|1\rangle\langle2| + i|2\rangle\langle1|),  \nonumber\\
\lambda_3 &= (|0\rangle\langle0| - |1\rangle\langle1|), \nonumber\\
\lambda_8 &= \frac{1}{\sqrt{3}}
\left(
|0\rangle\langle0| + |1\rangle\langle1| - 2|2\rangle\langle2|
\right).
\end{align}
Physically, the operators in the first three rows correspond to coherences within the three two-level subspaces spanned by $\{\ket{0},\ket{1}\}$, $\{\ket{0},\ket{2}\}$ and $\{\ket{1},\ket{2}\}$, respectively. The operators $\lambda_3$ and $\lambda_8$ are diagonal and respectively correspond to the difference between the population in the $\ket{0}$ and $\ket{1}$ levels, and the difference between the total population in the $\{\ket{0},\ket{1}\}$ subspace and twice the population in the third level. These two diagonal operators commute and constitute a Cartan sub-algebra~\cite{georgi2000lie} for ${\rm su}(3)$.

To describe the dynamics and observables corresponding to a collection of $N$ atoms, we introduce the collective Gell-Mann operators, defined as 
\begin{equation}
\label{N_Particle_gen}
\Lambda_k = \frac{1}{2}\sum_{j=1}^{N} \lambda_k^{(j)},
\end{equation}
where $k=1,\ldots,8$ and $\lambda_k^{(j)}$ is the $k$-th Gell-Mann matrix for atom $j$. Here the prefactor \(\frac{1}{2}\) is introduced to ensure the convention of SQL and HL introduced in Sec.~\ref{single_para_theory}.

In addition, for an $N$-qutrit system, we denote the total population difference between two states \(a\) and \(b\) by the collective operators \(\{\Lambda_{ab}\}\), where \(a,b \in \{0,1,2\}\). These operators are defined as
\begin{equation}
\label{pop_diff}
\Lambda_{ab} = \frac{1}{2}\sum_{k=1}^{N} \lambda_{ab}^{(k)},
\end{equation}
where the corresponding single-particle population imbalance operators are
\begin{equation}
\lambda_{ab} = \ket{a}\bra{a} - \ket{b}\bra{b}.
\end{equation}
It is worth noting that \(\lambda_{01}\) is identical to the standard single-particle Gell-Mann matrix \(\lambda_3\).

\subsection{ One-Axis Twist (OAT) model}
Dynamically generating metrologically useful states requires engineered Hamiltonians that can induce correlations between atoms. One such Hamiltonian is the one-axis twist Hamiltonian~\cite{kitagawa1993squeezed,begzjav2021squeezing} given by 
\begin{equation}
\label{Qubit_OAT}
    H_{\rm OAT}=\chi J_z^2.
\end{equation}
Here \(J_z=\frac{1}{2}\sum_k \sigma^k_z\) and \(\sigma^k_z\) is the usual Pauli-Z operator for the \(k-\)th atom, i.e., \(\sigma_z=\ket{0}\bra{0}-\ket{1}\bra{1}\). 
Starting from an initial coherent state of the form 
\begin{equation}
    \ket{\psi_{\rm in}}=\frac{1}{2^{N/2}}(\ket{0}+\ket{1})^{\otimes N},
\end{equation}
The OAT Hamiltonian generates entangling dynamics between the two levels of the atoms, leading to spin-squeezed states at short times and highly entangled non-Gaussian states at longer evolution times. Such states exhibit enhanced metrological sensitivity beyond the standard quantum limit for measuring collective rotations about specific axes on the Bloch sphere \cite{pezze2018quantum}. 

Inspired by this, we consider three qutrit extensions of the qubit OAT. In each case we start with a coherent state with equal superpositions of all three levels of the form
\begin{equation}
\label{initial_co}
    \ket{\psi_{\rm in}}=\frac{1}{3^{\frac{N}{2}}}(\ket{0}+\ket{1}+\ket{2})^{\otimes N}.
\end{equation}
Subsequently, we evolve the initial state under the Hamiltonian and characterize the metrological utility of the dynamically generated states.

\subsection{Qutrit One-Axis Twist (QOAT) model} 

The first model we discuss is the qutrit one-axis twist (QOAT) Hamiltonian, which is essentially the standard OAT Hamiltonian acting only in a 2-level subspace with the third level serving as a spectator level. The QOAT Hamiltonian is given by
\begin{equation}
H_{\rm QOAT} = \chi \Lambda_{01}^2 .
\label{H1}
\end{equation}
Essentially, Eq.~(\ref{H1}) can be expanded as
\begin{equation}
\label{H_QOAT}
H_{\rm QOAT} = \frac{\chi}{4}(\sum_{i}^N\lambda^{i}_{01}\lambda^i_{01} +\sum_{k\neq l} \lambda^{k}_{01} \lambda^l_{01}).
\end{equation}
In the case of the standard OAT, since \(\sigma^i_z \sigma^i_z=I\), the single particle contribution is just proportional to identity, whereas in the qutrit case, due to the presence of the third level, we have  
\begin{equation}
    \lambda_{01}\lambda_{01}=I-\ket{2}\bra{2}.
\end{equation}
This makes the single particle term depend on the population of the third level.

\subsection{Balanced One-Axis Twist model}
The second model we introduce is the balanced one-axis twist (BOAT) Hamiltonian~\cite{upadhyay2025scalable}. While QOAT performs entangling operations on one specific pair of internal levels, BOAT extends this by performing entangling operations on all three two-level subspaces. If the nonlinear interaction strength is identical across all subspaces, the resulting dynamics treat all three internal levels on an equal footing, leading to a balanced distribution of nonlinear interactions across the qutrit manifold. The BOAT Hamiltonian is given by
\begin{equation}
    H_{\rm BOAT}=\frac{\chi}{3}(\Lambda^2_{01}+\Lambda^2_{12}+\Lambda^2_{20}).
\end{equation}
Here \(\Lambda_{12}\) and \(\Lambda_{20}\) are the collective population difference operators defined in Eq.~(\ref{pop_diff}). As the three terms in $H_{\rm BOAT}$ commute,  BOAT can be realized starting from QOAT by successively implementing the latter Hamiltonian in the three different two-level subspaces. Since three QOAT operations are required to realise BOAT, the factor \(1/3\) is used to normalise the interaction strength so that the accumulated phase over a time $t$ can be compared to the QOAT Hamiltonian. Furthermore, BOAT only requires the capability to implement QOAT in a single two-level subspace. The remaining two terms can be obtained by sandwiching QOAT between appropriate $\pi$-pulses that swap the populations between various levels~\cite{upadhyay2025scalable}. Therefore, this model also maintains the third level as a spectator during the entangling dynamics. 

\subsection{Qutrit XY model}

The third model we consider is the qutrit XY (QXY) Hamiltonian. The motivation for this model comes from spin-exchange interactions commonly encountered in cavity QED systems, which can be used to realize OAT type interactions~\cite{sorensen2002entangling,leroux2010implementation,kitagawa1993squeezed}. In the case of $N$ identical qubits coupled off-resonantly to a cavity, the effective atomic dynamics after adiabatically eliminating the cavity are governed by the Hamiltonian 
\begin{equation}
    H = \chi J_{+}J_{-},
\end{equation}
where \(J_{\pm}\) are the collective spin raising and lowering operators. This Hamiltonian essentially realizes an all-to-all coupled XY model~\cite{lee2014XY_model}. Using angular momentum algebra, we can rewrite this Hamiltonian as 
\begin{equation}
\label{OAT_H}
    H = \chi(J^2 - J_z^2 + J_z),
\end{equation}
which reveals the connection to the OAT Hamiltonian: The Hamiltonian is symmetric under exchange of atomic indices. If, in addition, the initial state is a fully symmetric state, then the dynamics is restricted to the \(J=N/2\) manifold. The \(J^2=J_x^2+J_y^2+J_z^2\) term becomes a constant of motion, leaving an unimportant linear term proportional to \(J_z\) together with the nonlinear OAT Hamiltonian \(J_z^2\).

Motivated by this connection to OAT, we generalise the spin-exchange interaction to the qutrit setting and define the qutrit XY (QXY) Hamiltonian as
\begin{equation}
    H_{\rm QXY} = \chi T_{+}T_{-},
\end{equation}
where
\begin{equation}
    T_{\pm} = \Lambda_1 \pm i\Lambda_2
\end{equation}
are the collective raising and lowering operators acting within the \(\ket{0}-\ket{1}\) subspace. Expanding the Hamiltonian in terms of the collective \(SU(3)\) generators gives
\begin{equation}
\label{QXY_H}
    H_{\rm QXY}=\chi ({K^2}-\Lambda_3^2+\Lambda_3).
\end{equation}
Here, \(K^2=\Lambda_1^2+\Lambda_2^2+\Lambda_3^2\) is the total angular momentum in the  \(\ket{0}-\ket{1}\) qubit subspace of the qutrit manifold. In contrast to the qubit case, \(K\) is not proportional to the identity operator due to the presence of the third qutrit level. Consequently, the Hamiltonian~\eqref{QXY_H} does not reduce to the QOAT Hamiltonian, unlike in the qubit case, where the spin-exchange Hamiltonian is equivalent to OAT up to a rotation, for appropriate initial conditions.

\subsection{Experimental realisation of the qutrit Hamiltonians}
The qutrit Hamiltonians proposed in this work can be experimentally realised on several experimental platforms. Here, we briefly discuss possible realisations in trapped-ion quantum hardware and cavity QED systems.

Reference~\cite{upadhyay2025scalable} suggested a number of potential implementation schemes for BOAT in trapped ion systems, which are equally applicable to the QOAT Hamiltonian as well. The general idea is to implement effective OAT interactions between any two selected levels of the qutrits using either the Mølmer--Sørensen gate~\cite{molmer1999multiparticle} or a light-shift gate~\cite{leibfried2003experimental}. Since the terms in the BOAT Hamiltonian commute, the BOAT model can be realised by sequentially applying OAT interaction between all pairs of levels. An alternate way to realise the BOAT model is to implement OAT between a fixed pair of levels and sequentially swap the levels. We note, however, that the effective nonlinear strength \(\chi\) of twisting dynamics realized in trapped ions typically scales as \(1/N\), where \(N\) is the number of ions. Accordingly, the corresponding timescales scale as $t\to Nt$ with system size. 

For neutral atoms in cavities, OAT can be implemented via a cavity feedback mechanism, as developed in Ref.~\cite{schleier2010squeezing} and implemented in Ref.~\cite{leroux2010implementation}. In this scheme, the steady-state intracavity photon number depends on the collective population difference of the atoms in the cavity. This results in an AC Stark shift proportional to the population difference, thereby generating a nonlinear OAT-type interaction between the two levels. By introducing an appropriate spectator level that can be coherently initialized with non-zero population, this scheme can be naturally extended to realise QOAT and BOAT interactions inside cavity. The QXY model can similarly be realized by combining existing demonstrations of spin-exchange interactions between qubits in cavities with the coherent initialization of a third spectator level \cite{sorensen2002entangling,luo2025hamiltonian}.

\section{Results: Single parameter estimation}
\label{single_para}
In this section, we present our results for single-parameter estimation. To quantify the metrological utility of the dynamically generated states under our models, we compute the largest eigenvalue of the covariance matrix \(\Sigma\), which determines the maximum achievable QFI. We also discuss the structure of the optimal generators. Finally, we investigate the scaling behaviour of the remaining eigenvalues of \(\Sigma\), providing insight into the occurrence of multiple sensitive directions in operator space.

\subsection{Maximum Quantum Fisher Information}

\begin{figure} 
    \centering
    \includegraphics[width=1\linewidth]{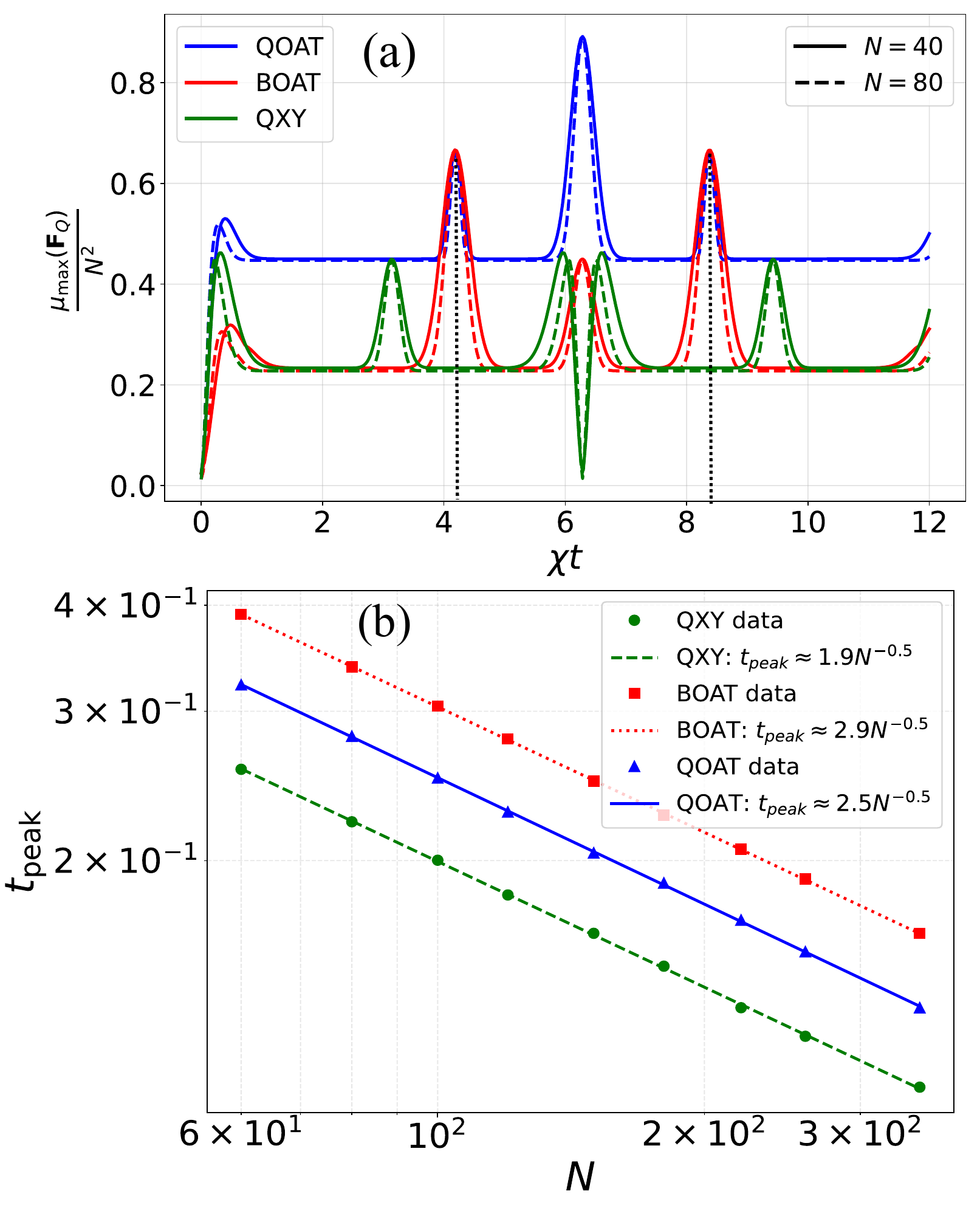}
    \caption{(a) Long time evolution of \(\mu_{\rm max}\) for QOAT (blue), BOAT (red), and QXY (green) model, solid lines and dashed lines correspond to \(N=40\) and \(N=80\) respectively. At \(\chi t = \frac{4\pi}{3}\) and \(\chi t = \frac{8\pi}{3}\), the black dotted vertical lines indicate the generation of the qutrit GHZ state.
    (b) Scaling of the \(t_{\rm peak}\) with system size \(N\) . For BOAT(points shown by square marker and fit line shown by red dotted line ), QOAT(points shown by triangular markers and fit line shown by blue solid line) and QXY(points shown by circular marker and fit shown by green dashed line) is respectively a \(2.9N^{-0.5},2.5N^{-0.5}\) and \(1.9N^{-0.5}\).}
    \label{fig:QFIM_long time}
\end{figure}

Fig.~\ref{fig:QFIM_long time}(a) shows the maximum eigenvalue of the covariance matrix \((\mu_{\rm max})\) normalised by \(N^2\) as a function of time for the three qutrit models considered in this work and for two system sizes with \(N=40\) and \(N=80\) qutrits. Beyond the initial transients, the ratio \((\mu_{\rm max})/N^2\) approximately converges to a system-size independent value in all three models, indicating the Heisenberg scaling of the maximal QFI. 

\subsubsection{Some non-classical states of interest}

The BOAT model has been previously studied in the context of its ability to generate qudit Greenberger-Horne-Zeilinger (GHZ) states~\cite{upadhyay2025scalable}. Here, the qutrit GHZ state is formed at $\chi t= 4\pi/3$ and is signalled by a peak in the maximal QFI~\footnote{The characteristic times observed here are fully consistent with Ref.~\cite{upadhyay2025scalable} after accounting for differences in numerical factors in the definition of $\chi$ in the two works.}. The QOAT Hamiltonian also realizes a qutrit GHZ state at the same time and its peak coincides with that in the BOAT model. Subsequently, the QOAT model achieves a larger peak QFI at $\chi t = 2\pi$. This state can be understood by interpreting the time evolution as OAT dynamics in the $\ket{0}-\ket{1}$ subspace with an effective number of qubits given by $N_{\rm eff} = N-N_2$, where $N_2$ is the population in $\ket{2}$: At \(\chi t=2\pi\) and in a given $N_2$ sector, the OAT-type interaction leads to a revival as the corresponding unitary becomes proportional to the identity in the $\ket{0}-\ket{1}$ subspace. However, when \(N_{\mathrm{eff}}\) is odd, the unitary acquires a phase factor \(e^{-i\pi/2}\), which is an inconsequential global phase for qubits but has a nontrivial consequence in the qutrit case; the initial product state in Eq.~(\ref{initial_co}) is a superposition of different $N_2$ sectors with a spread of \(\Delta n\sim\sqrt{N}\). This coherently populates multiple even and odd \(N_{\mathrm{eff}}\) sectors in the $\ket{0}-\ket{1}$ subspace, which converts the revival phase into a relative phase between different population sectors. This results in a non-trivial entangled state with enhanced QFI at $\chi t = 2\pi$. In contrast to standard OAT with revival time $\chi t = 2\pi$, the revival time for QOAT is $\chi t = 8\pi$ as it takes $4$ OAT cycles to bring the phase factor of the odd $N_{\rm eff}$ sectors to unity.  

Notably, the QOAT and QXY models lead to qualitatively different dynamics. In particular, the QXY model appears to neither generate a qutrit GHZ state nor the nontrivial high QFI state at \(\chi t=2\pi\), in contrast to the QOAT model. This suggests that, unlike in qubit systems, where XY-type interactions are closely connected to OAT dynamics, the QOAT and QXY models produce fundamentally different correlation structures in qutrit systems. 

\subsubsection{First peak and scaling}

Fig.~\ref{fig:QFIM_long time}(a) shows that, while the time required to reach the subsequent peaks, including the qutrit GHZ peak, remains independent of the system size $N$, the time required to reach the first peak decreases with $N$. We denote the time at which the maximum eigenvalue of $\Sigma$ attains its first maximum as $t_{\mathrm{peak}}$. Fig.~\ref{fig:QFIM_long time}(b) shows that $t_{\rm peak}\propto N^{-1/2}$ all the three models considered, analogous to the qubit OAT setting. 

The observed $N^{-1/2}$ scaling admits a qualitative physical interpretation based on the phase-shearing mechanism familiar from nonlinear collective dynamics~\cite{kitagawa1993squeezed,pezze2009entanglement}. The initial coherent qutrit product state is a superposition of collective population configurations centered around the balanced population $(N/3,N/3,N/3)$, with a characteristic population spread $\Delta n\sim\sqrt{N}$ arising from the multinomial statistics. During subsequent evolution under Hamiltonians that solely depend on population differences between levels, different population configurations accumulate different dynamical phases, causing neighbouring configurations to acquire a relative phase at a characteristic rate $\sim\chi\Delta n\sim\chi\sqrt{N}$. Consequently, the accumulated phase spread across the occupied population distribution grows as $\Delta\phi\sim\chi\sqrt{N}\,t$. The first significant deformation of the collective wavepacket, corresponding to the onset of strong multipartite correlations and the first peak of the largest  eigenvalue of $\Sigma$, occurs once this phase spread becomes of order unity, i.e., $\chi\sqrt{N}\,t_{\mathrm{peak}}\sim1$, which immediately yields
$t_{\mathrm{peak}}\propto(\chi\sqrt{N})^{-1}\propto N^{-1/2}
$, in qualitative agreement with the numerical scaling observed for all three models.

Owing to the inverse scaling with $N$, $t_{\mathrm{peak}}$ is experimentally attractive for metrological applications, since shorter evolution times are preferred to minimize the impact of background decoherence processes. We note that, even in the case of trapped ion implementations where $\chi\propto 1/N$, the scaling of \(t_\mathrm{peak}\) remains sublinear, i.e. \((t_{\rm peak}\propto N^{1/2})\), while the subsequent peaks show linear scaling with $N$. Hence, for the remainder of this work, we primarily analyze the metrological performance of the three models at $t_{\mathrm{peak}}$. 

\subsection{Structure of the optimal generators}
\begin{figure}
    \centering  
    \includegraphics[width=1\linewidth]{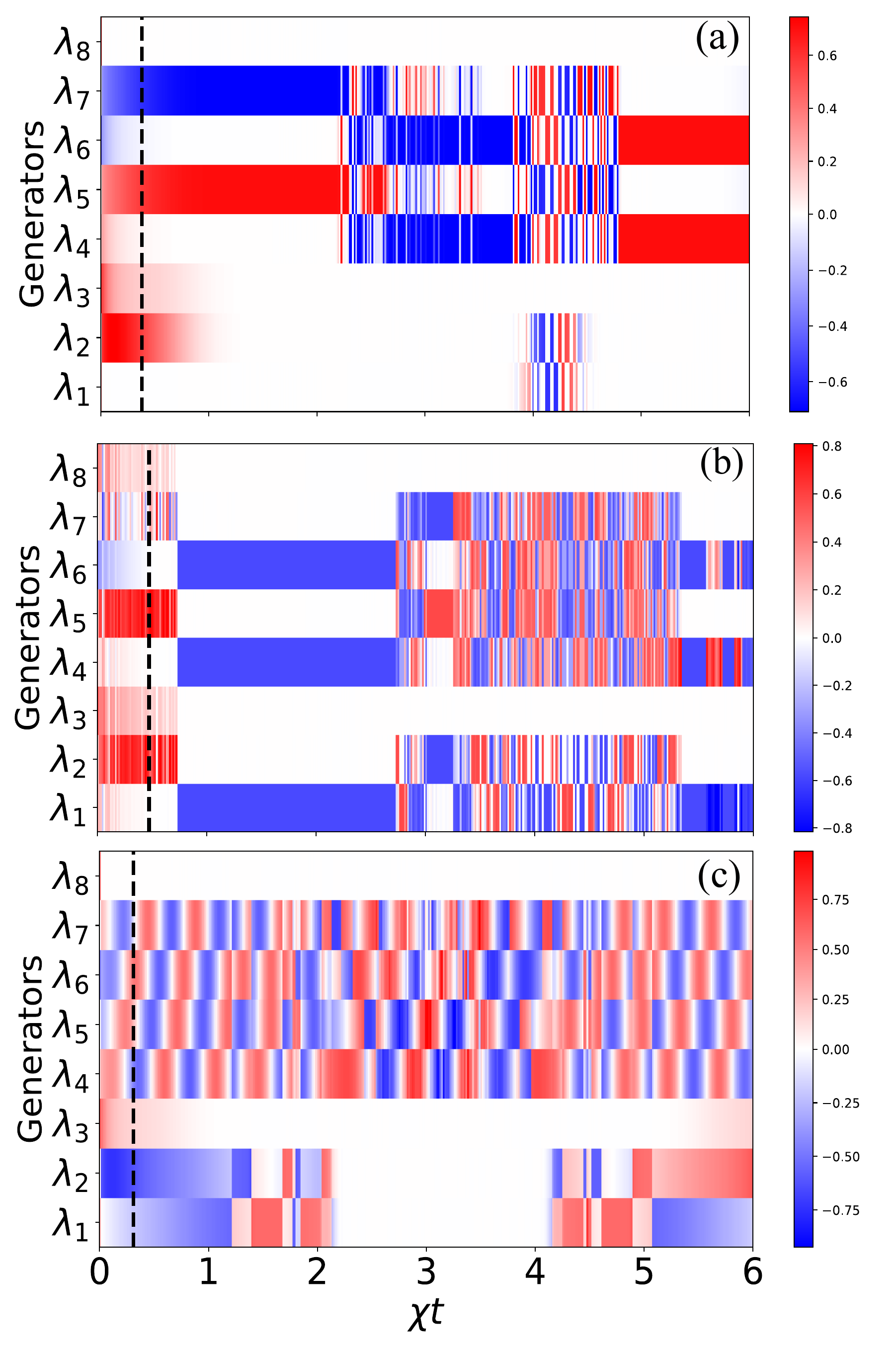}
    \caption{Contribution of the Gell-Mann operators in the optimal generator \(G_{\mathrm{opt}}\) for (a) QOAT \((N=40)\),
    (b)BOAT\((N=40)\) and (c) QXY \((N=40)\). The vertical black dashed lines indicate the \(t_{\rm peak}\) for each model. At \(t_{\rm peak}\), for the QOAT and QXY models, the optimal generator has no contribution from \(\lambda_8\). However, there is very little contribution from \(\lambda_8\) for the BOAT model.}
    \label{fig:G_opt}  
\end{figure}

While the maximum eigenvalue of $\Sigma$ reflects the best achievable QFI for a given state, the corresponding eigenvector reveals the encoding direction that achieves this optimal sensitivity. Fig.~\ref{fig:G_opt} shows the contribution of the eight Gell-Mann operators [Eq.~(\ref{Gell-Mann})] to the optimal generators in the QOAT, BOAT and QXY models as a function of evolution time. The time corresponding to $t_{\rm peak}$ is also indicated in each case. The rapid fluctuations in the signs of the coefficients in certain intervals of time can be attributed to an ambiguous global factor of $\pm 1$ in defining the optimal generators at any given time. In addition, the BOAT model has two degenerate optimal generators at $t_{\rm peak}$, discussed further in Sec.~\ref{sec:heisenberg_mult}, which provides greater freedom in encoding direction, but leads to ambiguity when visualizing a single optimal generator. 

.
\begin{figure}
    \centering
    \includegraphics[width=1\linewidth]{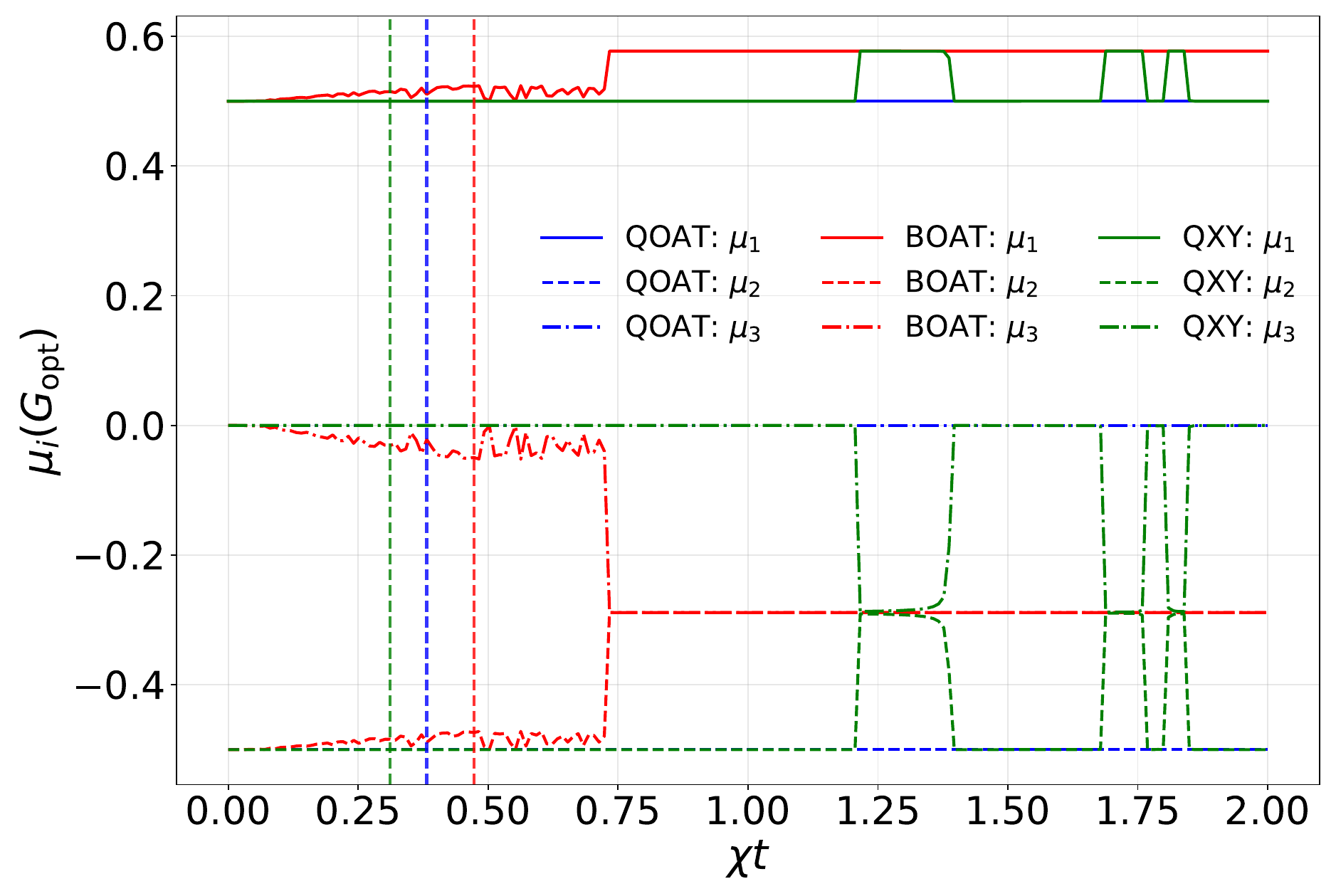}
    \caption{The eigenvalues of the optimal generator for QOAT (Blue), BOAT(Red) and QXY(Green) models have been shown for short time duration. The vertical dashed lines mark the \(t_{\rm peak}\) for the respective models. At \(t_{\rm peak}\) for the QOAT and QXY model, the eigen spectrum is strictly \(\frac{1}{2}\{1,-1,0\}\). Whereas for the BOAT model, the spectrum is close to \(\frac{1}{2}\{1,-1,0\}\). }
    \label{fig:G_opt_eigenvalue}
\end{figure}

For the QOAT model, the optimal generator at \(t_{\rm {peak}}\) has dominant contributions arising from \(\lambda_2\), \(\lambda_5\), and \(\lambda_7\), while the contributions from \(\lambda_3\),\(\lambda_4\) and \(\lambda_6\) are comparatively small. Notably, the contribution from \(\lambda_8\) is exactly zero within numerical precision. For the QXY model, the optimal generator at \(t_{peak} \) has major contributions from \(\lambda_2\), \(\lambda_6\) and \(\lambda_4\), relatively small contributions from the rest of the operators, and no contribution from \(\lambda_8\), similar to QOAT. Finally, in the case of BOAT model, the optimal generator at \(t_{\rm peak}\) that is shown here is dominated by \(\lambda_2\) and \(\lambda_5\), with small but finite contributions from \(\lambda_8\), \(\lambda_3\) and \(\lambda_7\). These observations about the structures of the optimal generators can guide experiments in identifying protocols for parameter encoding in order to optimize the sensitivity. 

In practice, parameter encoding in the optimal direction will be greatly simplified if the optimal generators can be mapped to commonly available qutrit operations via a unitary transformation. For this to be possible, the optimal generator must share the same eigenspectrum as one of the commonly available qutrit operators. In that case, the parameter encoding can be done with the latter operator, which can then be unitarily transformed into the optimal generator. The commonly available qutrit operations often involve driving transitions and realizing $Z$-type gates between pairs of levels. The operators driving transitions correspond to $\lambda_j/2, j=1,2,4,5,6,7$ while $Z$-type gates require Hamiltonians that contain population difference operators, e.g., $\lambda_3/2$. All of these generators have the eigenspectrum $\{-1/2,0,1/2\}$ and hence optimal generators can be unitarily mapped to these common operations if they share the same eigenspectrum. To see if this is the case, in Fig.~\ref{fig:G_opt_eigenvalue}, we plot the eigenvalues of the optimal generators as a function of time. For the QOAT and QXY models, we observe that the optimal generators have exactly the desired eigenspectrum at $t_{\rm peak}$, whereas for BOAT the eigenspectrum is approximately of the required form. In fact, we find that the optimal generator is twofold degenerate in the BOAT model at and around $t_{\rm peak}$, and that a suitable linear combination of the two generators can be found such that the resulting eigenspectrum is exactly $\{-1/2,0,1/2\}$.  Hence, in all three models, optimal encoding directions can in principle be accessed by sandwiching a parameter encoding involving one of the first $7$ Gell-Mann matrices in between an appropriate unitary transformation that connects the chosen encoding generator to the optimal generator.

\subsection{Near-Heisenberg scaling of multiple eigenvalues}
\label{sec:heisenberg_mult}

\begin{figure}
    \centering
    \includegraphics[width=1\linewidth]{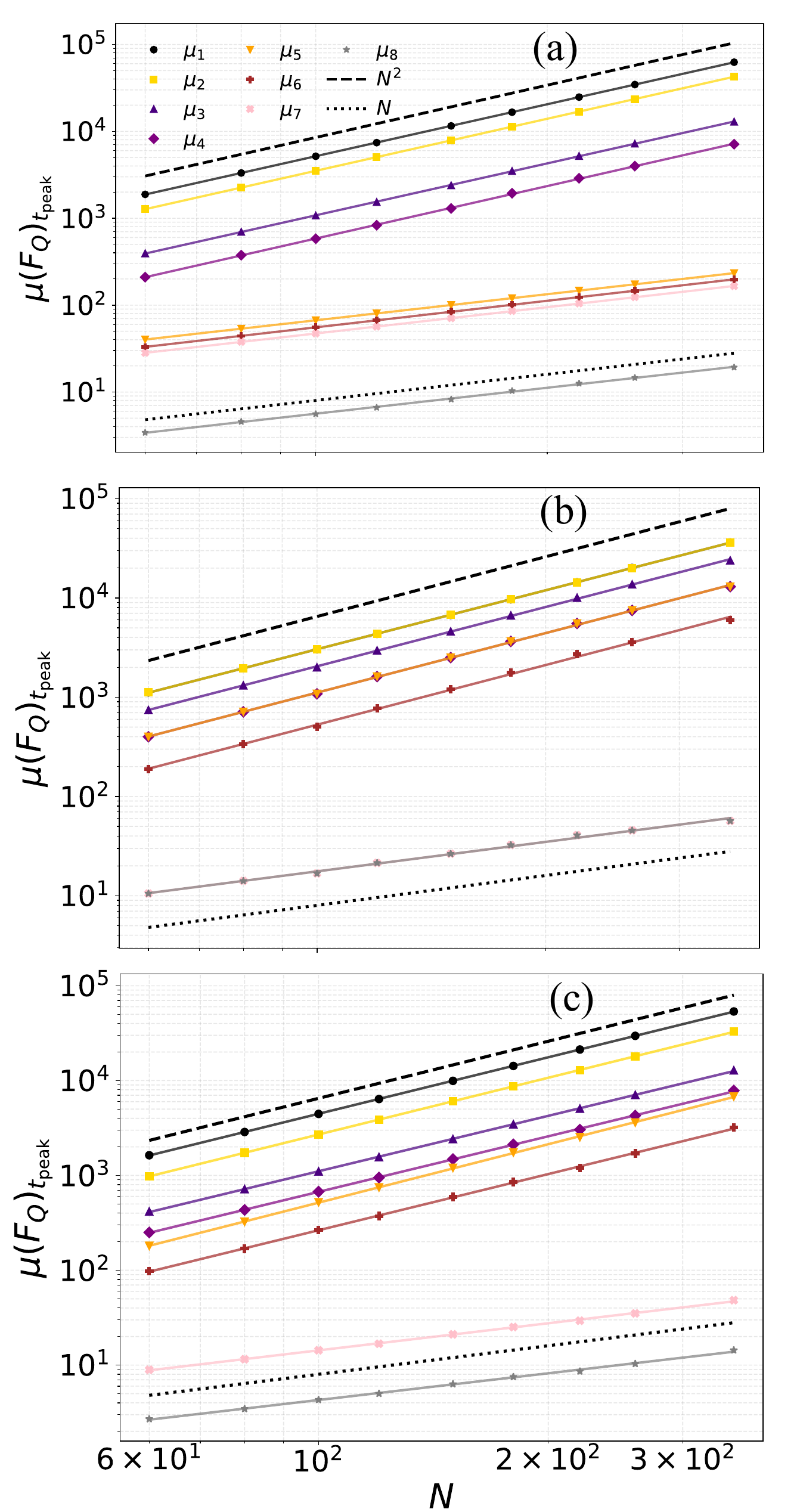}
    
    \caption{Scaling of eigenvalues of \(\Sigma\) matrix at \(t_{\rm peak}\) with system size \(N\) for (a)QOAT, (b) BOAT and (c) QXY models. The solid lines are linear fits, and the markers are the plot points. The slope of the fit lines gives the exponent of \(N\). QOAT has a $4$ orthogonal generators for which sensitivity scales approximately as Heisenberg scaling \((\propto N^2)\), whereas BOAT and QXY have $6$ such generators. The \(N^2\) and \(N\) scaling lines are shown as dashed and dotted black lines, respectively, in each plot.}
    \label{scaling_Cov}
\end{figure}

Beyond the maximal eigenvalue of $\Sigma$ and the associated optimal generators, the generated states may also be highly sensitive to parameter encodings using other generators, especially given the larger Hilbert space explored by qutrit systems compared to qubits. In Fig.~\ref{scaling_Cov}, we plot all eight eigenvalues of the \(\Sigma\) matrix at \(t_{\rm peak}\) as a function of system size \(N\) for the QOAT, BOAT and QXY models. In the QOAT model, the top four eigenvalues of \(\Sigma\) scale nearly as $N^2$, indicating near-Heisenberg scaling, whereas for the BOAT and QXY models, the top six eigenvalues exhibit near-Heisenberg scaling. This implies that, for the BOAT and QXY models, there exists only a \(2 \times 2\) subspace of generators whose sensitivity scales similar to the SQL (QFI $\propto N$), while for the QOAT model, the corresponding subspace is of size \(4 \times 4\).

Furthermore, at \(t_{\mathrm{peak}}\), the BOAT model exhibits three pairs of two-fold degeneracies in the eigenspectrum, namely between the first and second, the fourth and fifth and the last two eigenvalues. The first two pairs exhibit near-Heisenberg scaling, while the last one exhibits near-SQL scaling. The degeneracy between the two largest eigenvalues implies the existence of two orthogonal generators that possess identical optimal sensitivity.

These results indicate that qutrit ensembles offer great freedom in the choice of parameter encoding direction to achieve near-Heisenberg scaling of measurement precision, thus making them a versatile platform for quantum-enhanced sensing and metrology.

\section{Measurement protocols to saturate QFI}
\label{attainibility_QFI}

While the QFI provides the ultimate upper bound on the sensitivity achievable with a given quantum state, achieving this bound in practice requires identifying experimentally feasible protocols to implement measurements in an optimal basis. To this end, we consider two measurement protocols, compute the corresponding classical Fisher information (CFI), and compare it with the QFI. 
We assume that all atoms are first pumped into the state \(\ket{0}^{\otimes N}\). Subsequently, the initial product state $\ket{\psi_{\rm in}}$ given by Eq.~(\ref{initial_co}) is prepared by applying a unitary $U_{\rm prep}$ such that 
\begin{equation}
    \ket{\psi_{\rm in}} = U_{\rm prep} \ket{0}^{\otimes N}. 
\end{equation}
The unitary $U_{\rm prep}$ consists of two successive collective rotations, i.e., $U_{\rm prep} = U_2 U_1$, where $U_1=\exp({-i2\Lambda_2 \arcsin{\sqrt{\frac{2}{3}}}}) $ and $U_2=\exp({-i\Lambda_7\frac{\pi}{2}})$. The unitary $U_1$ coherently transfers two-thirds of the total population in $\ket{0}$ to level \(\ket{1}\). Subsequently, $U_2$ coherently transfers half of the population in \(\ket{1}\) to level \(\ket{2}\). 
Next, the initial state is acted upon by an entangling unitary $U_{\rm en}=e^{-iHt}$ to generate an entangled probe state,
\begin{equation}
\ket{\psi} = U_{\rm en} U_{\mathrm{prep}}\ket{0}^{\otimes N},
\end{equation}
where $H$ is one of the three twisting Hamiltonians considered in this work. The unknown parameter $\phi$ is then encoded through a unitary
\begin{equation}
U_{\phi} = e^{-i \phi G_{\mathrm{opt}}}.
\end{equation}
To achieve the maximum possible QFI, here we use the optimal generator for the corresponding model and evolution time to encode the unknown phase. 

The sensor operation is finally completed by performing a measurement and estimating the unknown phase. We assume that the experimentally accessible measurement is a readout of populations in the three qutrit levels, which is typically the case in several platforms. However, prior to this readout, we consider applying a decoding unitary $U_{\rm dec} = U_{\rm prep}^\dagger U_3$ that effectively transforms the measurement basis. The unitary $U_3$ will be discussed shortly below. Thus, the final state prior to population readout is 
\begin{equation}
\ket{\psi_{\mathrm{f}}}
=  U_{\rm prep}^\dagger U_{3} U_{\phi} U_{\rm en} U_{\mathrm{prep}}\ket{0}^{\otimes N}.
\end{equation}
For \(N\) qutrits, a single measurement outcome is specified by the population configuration \(\mathbf{K}=(N_0, N_1, N_2)\), where \(N_0+N_1+N_2=N\). The corresponding probabilities become 
\begin{equation}
    P_{\mathbf{K}}=|\langle N_0,N_1,N_2\ket{\psi_f}|^2.
\end{equation}
Repeating the experiment multiple times allows one to estimate the unknown phase $\phi$, e.g., by performing a maximum likelihood estimation, which is an optimal strategy that saturates the classical Fisher information (CFI) in the limit of a large number of measurement repetitions. Hence, we assess the sensor performance by computing the CFI, which is given by
\begin{equation}
\label{CFI}
F_C(\phi)=\sum_{\mathbf{K}} \frac{1}{P_{\mathbf{K}}(\phi)}
\left(\frac{\partial P_{\mathbf{K}}(\phi)}{\partial\phi}\right)^2,
\end{equation}
and compare it against the QFI of the entangled probe state $\ket{\psi}$.

Here, we consider two different choices for $U_3$. For the  \textit{identity} protocol, we choose
\begin{equation}
U_3 = I.
\end{equation}
The second choice corresponds to a \textit{time-reversal} protocol, where 
\begin{equation}
U_3 = U_H^{\dagger}.
\end{equation}
In this case, the nonlinear dynamics used to prepare the entangled probe state is reversed after parameter encoding. In practice, time-reversal of twisting dynamics can be achieved on multiple platforms by reversing the sign of the interaction, e.g., by changing a detuning parameter~\cite{garttner2017measuring}. The time-reversal protocol is widely used in nonlinear interferometry and spin-squeezing protocols~\cite{macri2016loschmidt,colombo2022time,davis2016approaching,linnemann2016quantum}.
Furthermore, in the case of zero operating point, i.e. when the true value of the parameter $\phi=0$, the time-reversal protocol is guaranteed to saturate the QFI since one of the measurement basis realized by this protocol corresponds to a projector onto the initial probe state itself. This latter property is a sufficient condition to saturate the QFI~\cite{humphreys2013quantum}.

\begin{figure}
    \centering
    \includegraphics[width=1\linewidth]{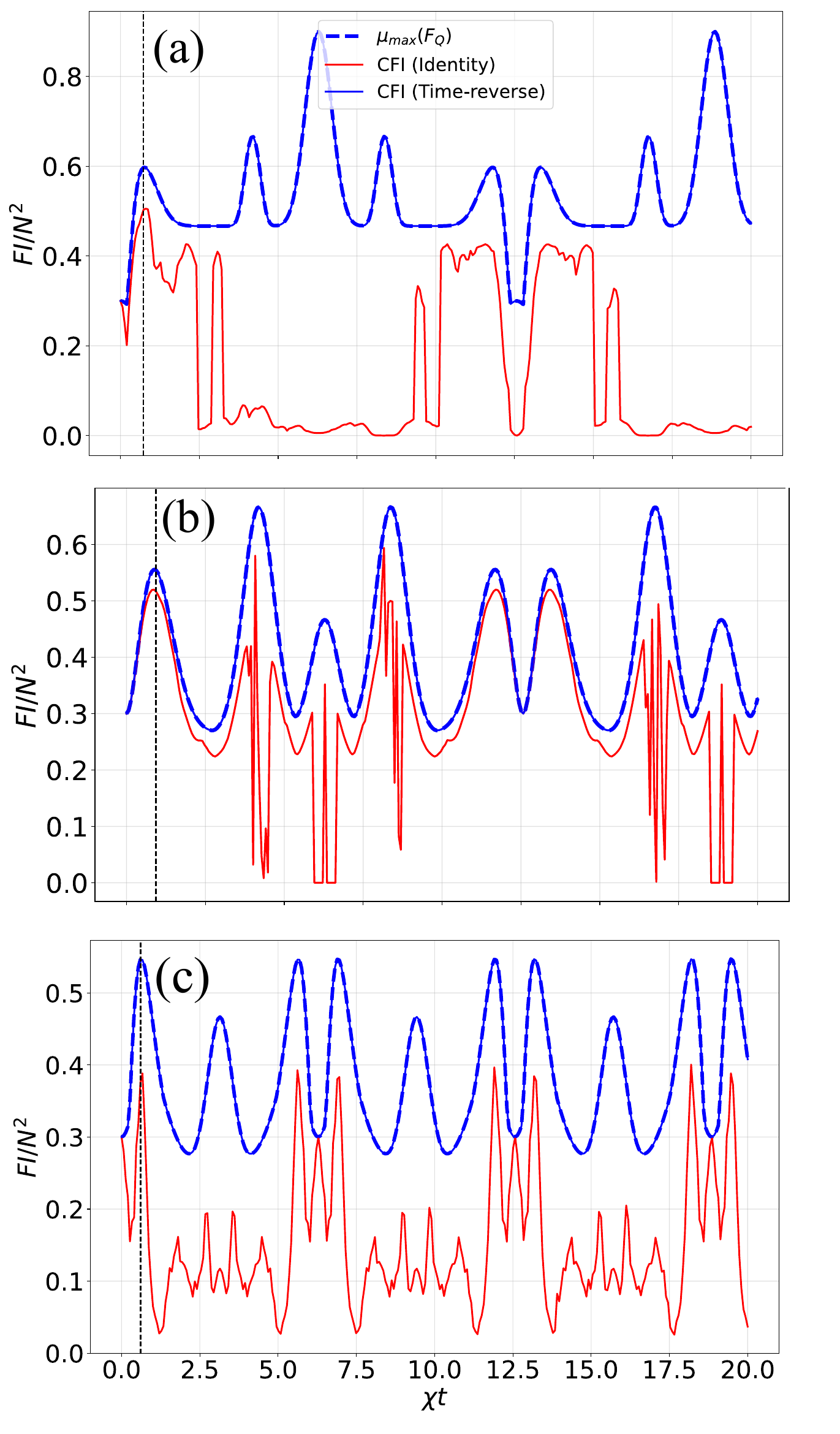}
    \caption{Comparison of the classical Fisher information (CFI) obtained using the time-reversal (solid blue line) and identity (solid red line) protocols with the maximum quantum Fisher information, $\mu_{\rm max}(F_Q)$ (blue dashed), for $N=10$. Results are shown for (a) the QOAT model, (b) the BOAT model, and (c) the QXY model. The vertical black dashed line marks the \(t_{\rm peak}\) for each model. The CFI obtained under the time-reversal protocol saturates the optimal QFI for all models, whereas under the identity protocol it does not. The same measurement basis also saturates the QFI corresponding to any other encoding direction.}
    \label{CFI_QFI}  
\end{figure}

In Fig.~\ref{CFI_QFI}, we compare the CFI defined in Eq.~(\ref{CFI}) obtained under the identity and time-reversal protocol with the QFI for the QOAT, BOAT and QXY models. As expected, the time-reversal protocol saturates the QFI across all the models and for all evolution times. However, the identity protocol fails to saturate the QFI over the dynamics, whereas in the QOAT and BOAT models, it approaches the QFI at short times.

In situations where it is challenging to reverse the sign of the Hamiltonian, another protocol can be considered in which the system is allowed to evolve further over time after the parameter is encoded. This is the \textit{time-forward protocol}, which is at the basis of recent studies of cyclic non-linear interferometry \cite{liu2023cyclic}. The corresponding unitary is given by
\begin{equation}
    U_3=e^{-iHt_2}
\end{equation}
 where \(t_2\) is the additional evolution time. The value of \(t_2\) can be optimised to maximise the resulting CFI. For the Hamiltonians considered in this work, the unitary dynamics are periodic. Specifically, the QOAT, BOAT, and QXY Hamiltonians satisfy \(e^{-iHT}=I\) (up to an overall global phase) for revival times corresponding to the dimensionless periods \(\chi T=8\pi\), \(4\pi\), and \(2\pi\), respectively. Expressing \(t_2=T-t'\) where \(t'\in [0,T]\), the corresponding unitary becomes $U_3=e^{-iHT} e^{iHt'}$. For \(t'=t\), we thus have $U_3 = e^{iHt}=U_H^\dagger$, which corresponds to the time-reversal protocol. Since the latter already saturates the QFI at \(\phi=0\), the time-forward protocol with $t_2=T-t$ is also optimal and can be implemented in situations where reversing the sign of the Hamiltonian is challenging.

Although we only show saturation of the QFI for the largest eigenvalue of \(\Sigma\) and its corresponding generator in Fig.~\ref{CFI_QFI}, the same time-reversal protocol is guaranteed to saturate the QFI for any encoding direction provided $\phi=0$ and can thus be used in conjunction with parameter encoding using any eigenvector of $\Sigma$.

\section{Some practical considerations}
\label{prac_con}
In this section, we examine the QOAT model to discuss the ambiguity in phase estimation at \(\phi=0\) and provide a way to circumvent it. Also, we discuss the effects of decoherence and provide a protocol to mitigate them. 
\subsection{Unambiguous phase estimation}
\label{unambiguous_phase}

\begin{figure}
    \centering
    \includegraphics[width=1\linewidth]{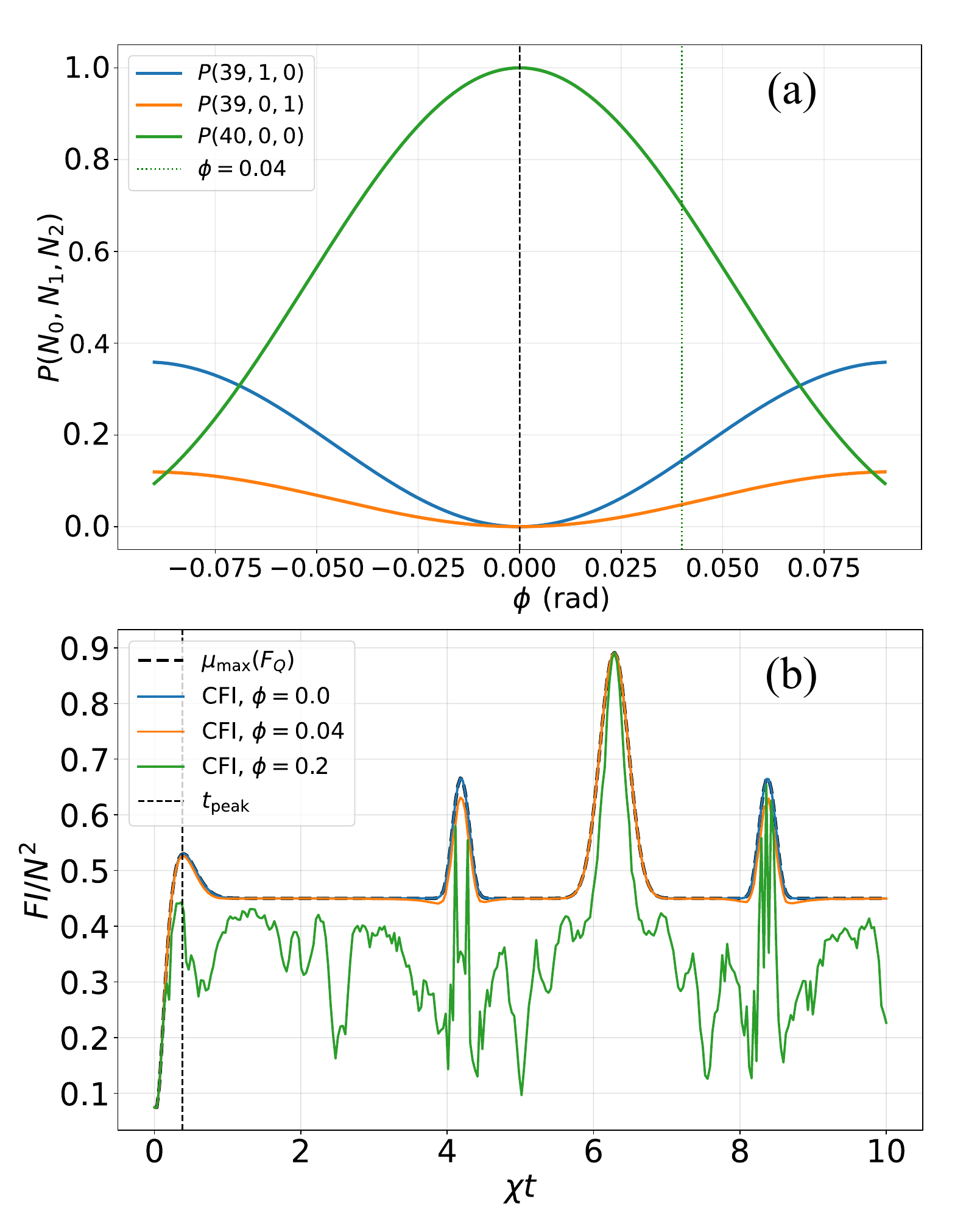}
    \caption{Unambiguous phase estimation at non-zero operating point for QOAT model with N=40. (a) probability distributions of the three most probable outcomes \(P(40,0,0), P(39,0,1)\) and \(P(39,1,0)\) are denoted by green, orange and blue solid lines, respectively. The distributions are symmetric about zero. The green dashed vertical line denotes \(\phi=0.04\), around which the probability distributions vary asymmetrically. (b) Optimal QFI (black dashed) and CFI obtained under time-reversal protocol for \(\phi=0\) (blue), \(\phi=0.04\) (orange) and  \(\phi=0.2\) (green). The vertical dashed line indicates the \(t_{\mathrm{peak}}\). For a non-zero operating point near zero, the CFI almost matches the QFI. Remarkably, at \(\chi t=2\pi\), the attainability of the QFI is robust for both the operating points.} 
\label{CFI_arbitary phi}
\end{figure}

The CFI of the time-reversal protocol saturates the QFI when the sensor operating point is at $\phi = 0$. However, as illustrated in Fig.~\ref{CFI_arbitary phi}(a), the probability distribution of outcomes is symmetric (even) around this point. In other words, the point $\phi=0$ is at the boundary of the domain of unambiguous estimation, which requires us to additionally know the sign of the unknown phase to estimate it unambiguously. A simple way to circumvent this ambiguity is to fix the operating point at a non-zero value. In Fig.~\ref{CFI_arbitary phi}(a), the green dotted line marks the operating point $\phi=0.04$. Around this point, the probabilities of measurement outcomes vary asymmetrically, thereby enabling the sensor to discriminate between small positive and negative deviations from this bias phase. The tradeoff, however, is that the CFI of the time-reversal protocol no longer exactly saturates the QFI. In Fig.~\ref{CFI_arbitary phi}(b), we plot the CFI of this protocol as a function of forward evolution time at the operating points $\phi=0.04$ and $\phi=0.2$ for a sensor with $N=40$ qutrits. We observe that at $\phi=0.04$, the CFI is very close to the QFI and hence operating with a small bias phase can be a practical strategy to overcome estimation unambiguity while maintaining high precision. Interestingly, at $\chi t = 2\pi$, the CFI saturates the QFI for both the operating points, which suggests the possibility of additional symmetries for the sensor state and measurement at this operation time ~\cite{miyazaki2022imaginarity}. We leave a detailed analysis of this feature to future work.

\subsection{Effect of decoherence}
\label{Dephasing}

Thus far, our analysis has assumed ideal unitary dynamics. In realistic implementations, however, decoherence is unavoidable and can significantly degrade metrological performance. Here, we investigate the effects of local decay and dephasing on the time-reversal protocol in the QOAT model. We evaluate the CFI for the measurement protocol described in Sec.~\ref{attainibility_QFI} and explore the possibility of mitigating the impact of decoherence processes by optimising the sensor operating point.
\begin{figure}
    \centering
    \includegraphics[width=1\linewidth]{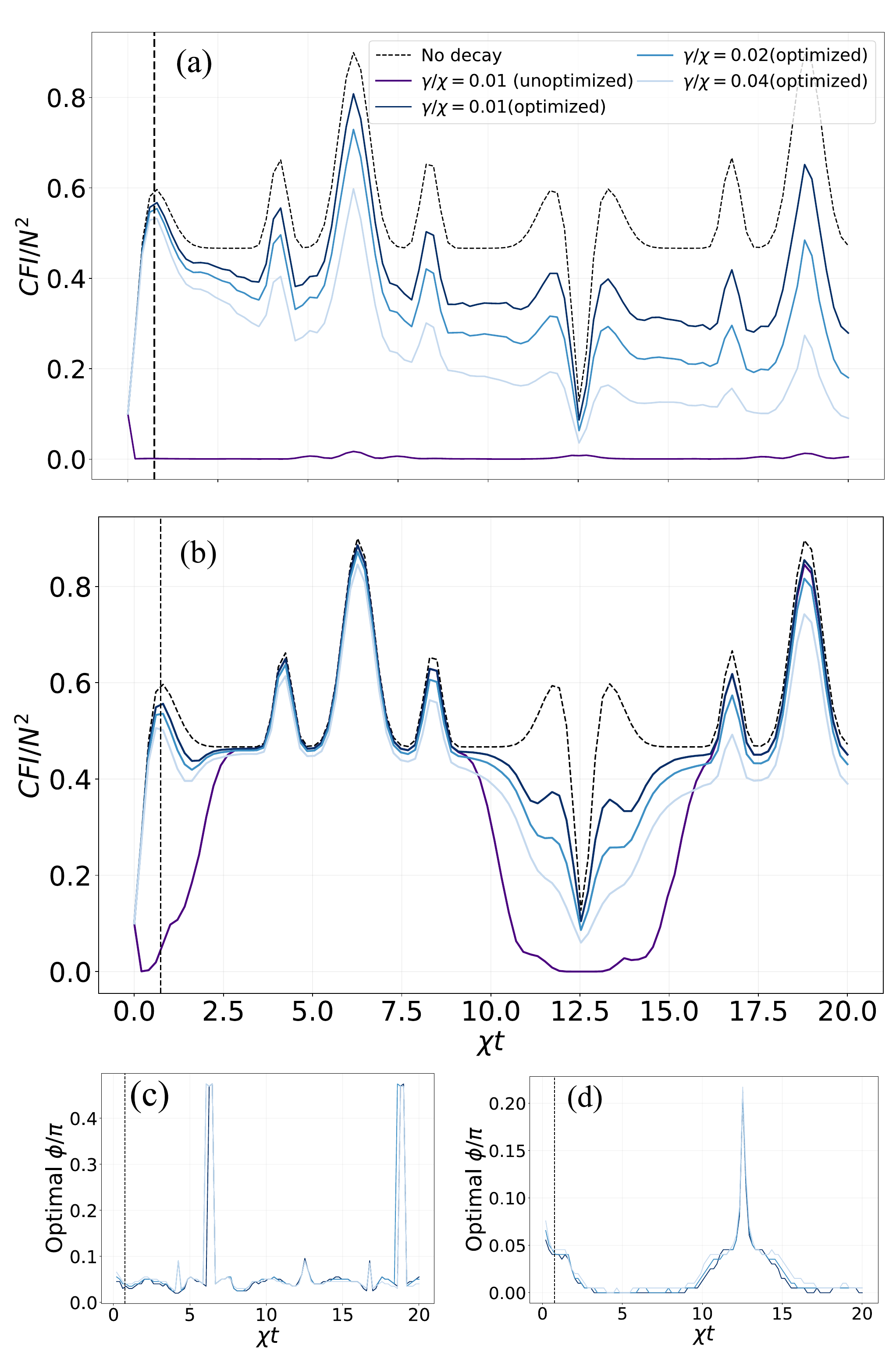}
\caption{Comparison of the conventional time-reversal protocol at zero operating point (solid violet line), and the optimised operating point time-reversal protocol (different Shades of blue ) with the no decoherence case (black dashed line). The classical Fisher information (CFI) is computed for the QOAT model with \(N=10\) under (a) local decay and (b) local dephasing for dissipation strengths \(\gamma/\chi=0.01\), \(0.02\), and \(0.04\) for both local decay and dephasing. (c) and (d) show the optimal operating points over time. The vertical black dashed line indicates the \(t_{\mathrm{peak}}\). At \(t_{\rm{peak}}\), CFI under the optimised time reversal protocol almost saturates QFI, whereas the conventional time-reversal protocol fails to saturate.}
    \label{fig:decoherence}
\end{figure}
To describe the dissipative dynamics of the QOAT model, we employ the Lindblad master equation given by
\begin{equation}
\label{eqn:master_eqn}
\frac{d\rho}{dt}
=
-\frac{i}{\hbar}[H,\rho]
+
\sum_k \left(
L_k \rho L_k^\dagger
-
\frac{1}{2} \{L_k^\dagger L_k, \rho\}
\right).
\end{equation}
Here, $L_k$ are local jump operators acting on the $k$th qutrit. In the case of local decay, we take 
\begin{equation}
    L_k
    =
    \sqrt{\gamma}\,\sigma_{01}^{(k)},
\end{equation}
where \(\sigma_{mn}=|m\rangle\langle n|\), (\(m,n\in\{0,1,2\}\)), while for local dephasing, we assume that 
\begin{equation}
    L_k
    =
    \sqrt{\gamma}\,
    (\sigma_{11}^{(k)}-\sigma_{00}^{(k)}).
\end{equation}
Such a noise model, where the local decoherence processes act only on the lower two levels is relevant, e.g., in the situation where the dominant source of decoherence is off-resonant scattering from the same lasers that drive the twisting dynamics in the $\ket{0}-\ket{1}$ subspace.  

In writing the above, we have assumed that the local decoherence processes occur with identical rates on all the qutrits. Together with the all-to-all homogeneous pairwise coupling in the QOAT model, this assumption makes the master equation invariant under arbitrary relabeling of qutrit indices. This permutation symmetry enables efficient simulation of the dynamics in the presence of decoherence:

For a system of \(N\) identical \(d\)-level particles, the dimension of the Liouville space associated with the density matrix scales in general as \(d^{2N}\). However, by restricting the dynamics to the permutation-symmetric subspace, this scaling is reduced to \(\mathcal{O}(N^{(d^2-1)})\)~\cite{gegg2016efficient}. For the qutrit \((d=3)\) systems considered in this work, the effective dimension therefore scales as \(\mathcal{O}(N^8)\), allowing us to efficiently simulate open system dynamics with $N\sim 10$ qutrits. To perform these simulations, we developed a Python implementation based on the framework introduced in Ref.~\cite{gegg2016efficient}. The associated source code is publicly available~\cite{sayam_repo}.

We summarise the impact of local decay and decoherence on the CFI in Fig.~\ref{fig:decoherence}. The black dashed line in both panels (a) and (b) shows the CFI of the time-reversal protocol in the absence of any decoherence at zero operating point, i.e. for $\phi=0$, which coincides with the QFI as demonstrated in the previous section. The violet line in Fig.~\ref{fig:decoherence}(a) shows the CFI when the local decay rate is $\gamma/\chi =0.01$. The curve shows that for $\phi=0$, the CFI drops close to $0$ for any non-zero protocol time, revealing the extreme sensitivity of this protocol to local decay processes. On the other hand, we find that shifting the sensor operating point away from $\phi=0$ makes the time-reversal protocol robust to local decoherence. In Fig.~\ref{fig:decoherence}(a), we show the CFI optimized with respect to $\phi$ as a function of protocol time and for increasing decay strengths in various shades of blue. These curves demonstrate that, at $t=t_{\rm peak}$, near-optimal performance comparable to the decoherence-free case can be achieved even in the presence of finite local decay by operating the sensor at an appropriate point. Fig.~\ref{fig:decoherence} (c) shows the optimal phase as a function of time. These curves reveal that the optimal phase is largely independent of the value of $\gamma$ over the range of values considered here. 

In Fig.~\ref{fig:decoherence}(b), we study the performance of the CFI under local dephasing. Once again, the CFI at $\phi=0$ degrades significantly even at low dephasing rates. However, at longer protocol times $\chi t\gtrsim \pi$, the CFI remarkably recovers and approaches the QFI of the decoherence-free protocol. Further optimizing the sensor operating point restores near-optimal performance even at short protocol times $t\sim t_{\rm peak}$. The Fig.~\ref{fig:decoherence}(d) reveals that, similar to the case of local decay, the optimal sensor operating point is largely independent of the dephasing rate.     

We note that the specific form of the local decay and dephasing terms in an actual qutrit experiment can vary sensitively depending on the specifics of the implementation. However, the above analysis suggests that, for low decoherence rates, the impact of such processes can be mitigated by optimizing the sensor operating point. 

\section{Results: Multi-Parameter Estimation}
\label{MUL}

The presence of several directions with high sensitivity raises the question of whether it is possible to simultaneously encode multiple parameters along the different directions and estimate them with the precision indicated by the QFI matrix (QFIM). As discussed in Sec.~\ref{multi_comp}, the attainability of the multiparameter quantum Cramer-Rao bound, Eq.~(\ref{MQCRB}), depends on the compatibility of the encoding generators as captured by the mean Uhlmann curvature defined in Eq.~(\ref{UC}). Here, we restrict our attention to the zero operating point of the sensor and first investigate the compatibility of the encoding generators by studying the parameter $R$ defined in Eq.~(\ref{R_para}) at \(t_{\rm {peak}}\). Subsequently, we demonstrate that the BOAT model exhibits a stronger version of compatibility between two of the encoding generators, which is not possible in qubit ensembles sensing global collective rotations.

\subsection{Measurement Incompatibility at $t_{\rm peak}$}

\begin{figure}
    \centering
    \includegraphics[width=1\linewidth]{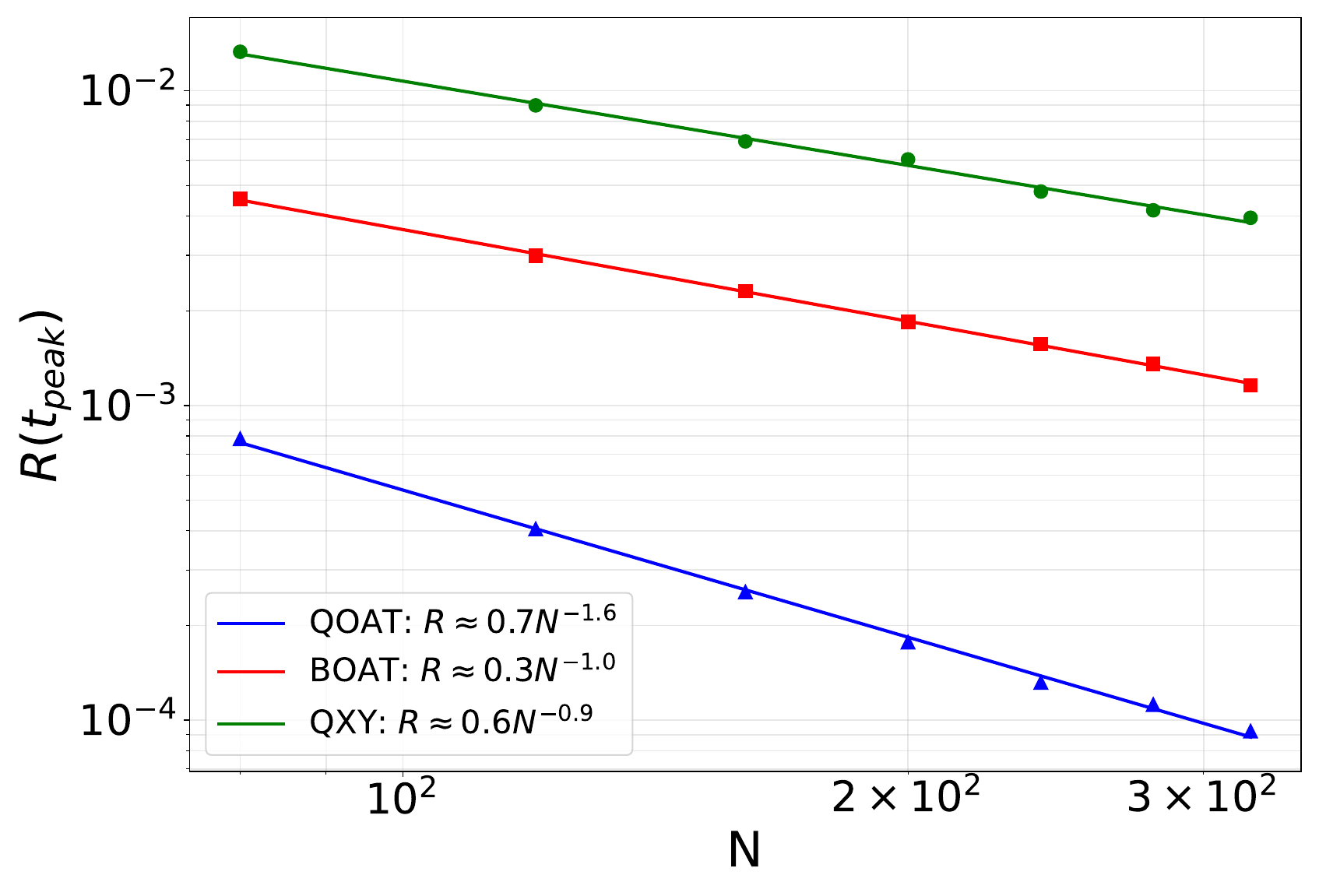}
\caption{The scaling of $R(t_{peak})$ with $N$ for the QOAT (blue), BOAT (red), and QXY (green) models. \(t_{\rm peak}\) for each model. As system size increases, the incompatibility between encoding generators decreases. Power-law fits give $R(t_{peak})\sim 0.7\,N^{-1.6}$ (QOAT), $R(t_{peak})\sim 0.3\,N^{-1}$ (BOAT) and $R(t_{peak})\sim 0.6\,N^{-0.9}$ (QXY).}
    \label{R}
\end{figure}

In order to study the measurement compatibility of the encoded parameters, we evolve the initial product state in Eq.~(\ref{initial_co}) under each of the three considered models and compute the QFIM [Eq.~(\ref{COV_mat})] and the Uhlmann curvature [Eq.~(\ref{UC})] for the state generated at $t_{\rm {peak}}$. Since we focus on the zero operating point of the sensor, the QFIM is equivalent to the covariance matrix $\Sigma$ defined in Eq.~(\ref{QFI}). We restrict our attention to the subspace of eigen-generators showing near-Heisenberg scaling of Fisher information at $t_{\rm peak}$. For this set of generators, we compute $R$ [Eq.~(\ref{R_para})] at $t_{\rm {peak}}$ and plot this quantity vs system size $N$ in Fig.~\ref{R}. 

 We observe that \(R(t_{\rm {peak}})\) scales as \(N^{-1.6}\), \(N^{-1}\) and \( N^{-0.9}\) for QOAT, BOAT and QXY models respectively, indicating strong suppression of multiparameter incompatibility at \(t_{\rm peak}\). Thus, according to Eq.~(\ref{eqn:hcrb_bounds}), the HCRB asymptotically approaches the QCRB with increasing $N$. Hence, increasing system size not only improves precision but also enhances compatibility among generators.

\subsection{Commuting eigen-generators in BOAT}

As reflected in the elements of the Uhlmann curvature matrix, Eq.~(\ref{UC}), the saturation of the multiparameter QCRB for parameters simultaneously encoded via two generators only requires the expectation value of their commutator to vanish when taken with respect to the probe state~\cite{baumgratz2016PRL}. However, $SU(3)$ sensors have the potential to host a stronger version of compatibility where the commutator itself vanishes.  Out of the three models considered here, we observe that the BOAT model possesses two eigen-generators that not only display near-Heisenberg scaling of QFI at $t_{\rm peak}$ but also perfectly commute over an extended time interval around this time, which we proceed to examine in further detail below. 

\begin{figure}
    \centering
    \includegraphics[width=1\linewidth]{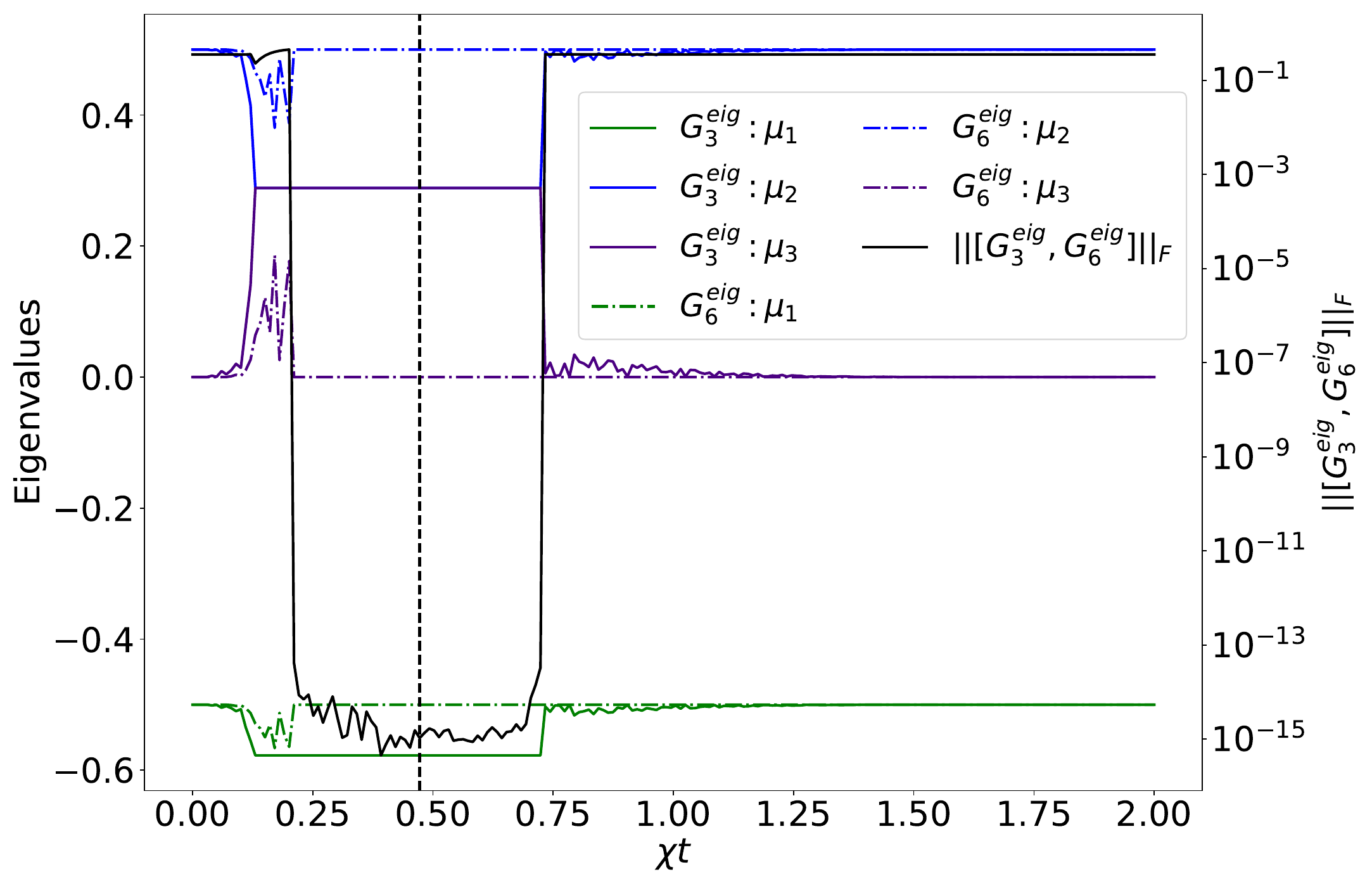}
    \caption {Dual-axis plot showing the eigenvalues and commutation properties of the third and sixth optimal generators in the BOAT model. The left y-axis displays the eigenvalues corresponding to the third optimal generator (solid line) and the sixth optimal generator (dash-dotted line). The right y-axis(\(\log\) scale) shows the Frobenius norm of their commutator. The vertical black dashed line indicates the  \(t_{\rm peak}\) of the BOAT dynamics. At \(t_{\rm peak}\), the Frobenius norm of the commutator is negligible, indicating that the two generators commute. Furthermore, the eigenvalue spectra of \(G^{\rm eig}_3\) and \(G^{\rm eig}_6\) becomes proportional to eigen spectrum of \(\lambda_8\) and \(\lambda_3\) respectively.}
    \label{G3G6}
\end{figure}

In Fig.~\ref{G3G6}, on the right $y$-axis, we plot the Frobenius norm of the commutator between \(G^{\rm eig}_3\)  and \(G^{\rm eig}_{6}\), which are the third and sixth eigenvectors of the \(\Sigma\) matrix, respectively, as a function of time. In an extended interval that also contains \(t_{\rm peak}\)   (vertical black dashed line), the norm is observed to be zero, demonstrating that both the generators commute. Notably, at \(t_{\rm peak}\) both the encoding generators \(G^{\rm eig}_3\) and \(G^{\rm eig}_{6}\) display near-Heisenberg scaling of the QFI with $N$. Hence, BOAT serves as a paradigm model that supports quantum-enhanced multiparemeter metrology on a pair of commuting eigen-generators.

The commuting eigen-generators can be traced back to the Cartan sub-algebra of $su(3)$, which contains two generators in contrast to just one for $su(2)$. 
On the left y-axis of Fig.~\ref{G3G6},  we plot the eigenvalues corresponding to \(G^{\rm eig}_3\) and \(G^{\rm eig}_6\). We find that at \(t_{\rm peak}\), \(G^{\rm eig}_3\) has eigenvalues \(\frac{1}{2\sqrt{3}}\{1,1,-2\}\) and \(G^{\rm eig}_6\) has eigenvalues \(\frac{1}{2}\{1,-1,0\}\). These eigenspectra coincide with those of the diagonal Gell-Mann operators \(\lambda_8\) and \(\lambda_3\) respectively. Moreover, since \(G^{\rm eig}_3\) and \(G^{\rm eig}_6\) commute, they can be simultaneously diagonalized. Therefore, there exists a unitary \(U\) such that
\begin{equation}
    \begin{pmatrix}
\lambda_8 \\
\lambda_3
    \end{pmatrix} = 
    U^\dagger
\begin{pmatrix}
G^{\rm eig}_3 \\
G^{\rm eig}_6
\end{pmatrix}
U.  
\end{equation}
The mapping of the encoding generators to \(\lambda_3\) and \(\lambda_8\) has an elegant interpretation in terms of simultaneous estimation of two phases. The generator \(\lambda_3\) encodes a relative phase between the states \(\ket{0}\) and \(\ket{1}\), while \(\lambda_8\) encodes the phase of \(\ket{2}\) relative to the \{\({\ket{0},\ket{1}}\)\} subspace. Hence, these phases correspond to the relative and global phases in an $SU(2)$ subspace of the full $SU(3)$ space of the qutrit. Furthermore, the commutativity of the two generators implies that the compatibility persists at any operating point of the sensor and is not limited to the zero operating point. 

\section{Summary and Outlook}
\label{sec:conc}

In this work, we investigated quantum-enhanced metrology with collective qutrit systems, extending beyond the conventional qubit-based paradigm for both single-parameter and multiparameter estimation. We considered three models that are generalizations of OAT dynamics to qutrit ensembles: the Qutrit One-Axis Twisting (QOAT), Balanced One-Axis Twisting (BOAT), and Qutrit XY (QXY) Hamiltonians and studied the metrological properties of the nonclassical states generated by their dynamics.

In particular, we focused on the quantum Fisher information (QFI) at the time \(t_{\rm peak}\), corresponding to the first maximum of the largest eigenvalue of the QFIM, which scales approximately as \(1/\sqrt{N}\). At this time, the optimal quantum Fisher information exhibits near-Heisenberg scaling with system size at $t_{\rm peak}$ for all three models. Moreover, the metrological enhancement is not restricted to a single encoding direction: At \(t_{\rm peak}\), the QOAT model possesses four orthogonal directions along which the state exhibits near-Heisenberg scaling of sensitivity, while the BOAT and QXY models possess as many as six directions with this property. Near the zero operating point of the sensor, the QFI can be attained through a time-reversal-based measurement protocol. Furthermore, we investigated the impact of local decay and dephasing on the sensor performance and found that their effects can be largely mitigated by optimizing the sensor operating point, yielding robust performance close to \(t_{\rm peak}\). We also note that operating at a non-zero operating point, preferably the optimised operating point under time-reversal, can circumvent the ambiguity in estimating \(\phi\) which arises at the zero operating point.

Subsequently, we investigated the precision with which multiple parameters can be simultaneously estimated using the states generated at \(t_{\rm {peak}}\). We found that the estimation incompatibility between parameters simultaneously encoded along different directions with near-Heisenberg scaling of sensitivity decreases with increasing system size for all three models. Furthermore, the BOAT model exhibits two encoding directions with near-Heisenberg scaling of sensitivity that exactly commute in the vicinity of $t_{\rm peak}$. This type of compatibility is absent in qubit sensors of collective $SU(2)$ rotations and is only possible due to the presence of two commuting generators in $su(3)$. This stronger version of the compatibility enables simultaneous estimation of two independent phases with perfect measurement compatibility at any sensor operating point. 

Overall, our results demonstrate that collective qutrit systems combine near-Heisenberg scaling of sensitivity, experimentally accessible measurement protocols, robustness against decoherence, unambiguous phase estimation at a non-zero operating point, and favourable compatibility properties for multiparameter estimation. These features establish collective qutrit dynamics as a versatile resource for quantum-enhanced sensing and precision measurement.

The present work opens multiple avenues for future exploration. While we have focused here on generalisations of OAT-type dynamics involving either only one nonlinear term or multiple commuting nonlinear terms in the Hamiltonian, it is reasonable to expect that Hamiltonians with two or more non-commuting terms can lead to faster generation of metrologically useful entanglement. This expectation stems from the established results in the qubit literature where the so-called twist-and-turn and two-axis countertwisting Hamiltonians lead to Heisenberg scaling of QFI on timescales $t\sim \ln(N)/N$ compared to $t\sim 1/\sqrt{N}$ in the OAT case~\cite{munoz2023phase,pezze2018quantum}. A second direction to explore is the use of systematic methods to identify optimal measurements that saturate the QFI while providing a large domain of unambiguous estimation around the operating point~\cite{vasilyev2024optimal}.  Thirdly, it will be interesting to identify applications that benefit from the strong compatibility of generators in the BOAT model, which enables simultaneous readout of a global and a relative phase in an appropriate two-level subspace with near-Heisenberg scaling of precision at any sensor operating point. Finally, a natural extension of this work is to investigate Bayesian approaches to developing optimal sensors that combine sensitivity with high dynamic range. Here, the collective qutrit Hamiltonians introduced in this work can be integrated as natively available gates in variational quantum circuits designed to identify optimal states and measurements that minimize a suitable cost function designed according to the sensing task at hand~\cite{kaubruegger2023optimal,marciniak2022optimal,kaubruegger2021quantum}.

\section*{Acknowledgments}

We thank Krishna Balaji for feedback on the manuscript. D.D, S.C. and A.S. acknowledge support by the Department of Science and Technology, Govt. of India through the INSPIRE Faculty Award (DST/INSPIRE/04/2023/001486), by the Anusandhan National Research Foundation (ANRF), Govt. of India through the Prime Minister’s Early Career Research Grant (PMECRG) (ANRF/ECRG/2024/001160/PMS), by IIT Madras through the New Faculty Initiation Grant (NFIG), and the support of the MPhasis F1 foundation to CQuiCC, IIT Madras. V.M. acknowledges support by ANRF, Govt. of India through Grant No. ANRF/ARGM/2025/002679/TS.

\bibliographystyle{apsrev4-2}
\bibliography{ref_2}
\end{document}